\documentclass{jfm}

\usepackage{graphicx}
\usepackage{natbib}
\usepackage{upmath}
\usepackage{subfig}
\usepackage{booktabs}
\usepackage{amssymb}%
\usepackage{amsmath}
\usepackage{pgfplots}
  
  \let\geq=\geqslant
\usepackage{amsbsy}
\allowdisplaybreaks

\newsavebox{\astrutbox}
\sbox{\astrutbox}{\rule[-5pt]{0pt}{20pt}}

\title[Kelvin ship wave patterns]{What is the apparent angle of a \\ Kelvin ship wave pattern?}

\author[R. Pethiyagoda, S. W. McCue and T. J. Moroney]%
{Ravindra Pethiyagoda$^1$,\ns
Scott W. McCue$^1$\thanks{Email address for correspondence: scott.mccue@qut.edu.au} \break
and Timothy J. Moroney$^1$}

\affiliation{$^1$Mathematical Sciences, Queensland University of Technology, QLD 4001, Australia}

\pubyear{2010}
\volume{650}
\pagerange{119--126}
\date{\today}
\begin{document}

\maketitle

\begin{abstract}
While the half-angle which encloses a Kelvin ship wave pattern is commonly accepted to be $19.47^\circ$, recent observations and calculations for sufficiently fast-moving ships suggest that the apparent wake angle decreases with ship speed. One explanation for this decrease in angle relies on the assumption that a ship cannot generate wavelengths much greater than its hull length. An alternative interpretation is that the wave pattern that is observed in practice is defined by the location of the highest peaks; for wakes created by sufficiently fast-moving objects, these highest peaks no longer lie on the outermost divergent waves, resulting in a smaller apparent angle. In this paper, we focus on the problems of free surface flow past a single submerged point source and past a submerged source doublet. In the linear version of these problems, we measure the apparent wake angle formed by the highest peaks, and observe the following three regimes: a small Froude number pattern, in which the divergent waves are not visible; standard wave patterns for which the maximum peaks occur on the outermost divergent waves; and a third regime in which the highest peaks form a V-shape with an angle much less than the Kelvin angle. For nonlinear flows, we demonstrate that nonlinearity has the effect of increasing the apparent wake angle so that some highly nonlinear solutions have apparent wake angles that are greater than Kelvin's angle. For large Froude numbers, the effect on apparent wake angle can be more dramatic, with the possibility of strong nonlinearity shifting the wave pattern from the third regime to the second.  We expect our nonlinear results will translate to other more complicated flow configurations, such as flow due to a steadily moving closed body such as a submarine.
\end{abstract}


\section{Introduction}
Kelvin ship wave patterns are comprised of divergent and transverse waves. For a steadily moving object in an infinitely deep inviscid fluid, the half-angle which encloses the ship wave pattern is widely accepted to be the Kelvin angle $19.47^\circ$, at least if the wave amplitude is small \citep{kelvin87}.  This angle can be determined by constructing geometric arguments with the dispersion relation \citep{lighthill}, by locating the caustics using the method of stationary phase \citep{ursell60}, or by applying geometric ray theory \citep{keller79}.  On the other hand, if the object is moving sufficiently fast, the angle that is actually observed in practice is less than the Kelvin angle, and in fact decreases with object speed \citep{rabaud13}. One explanation for these observations, proposed by \citet{rabaud13}, is that a ship cannot generate wavelengths much greater than its hull length \citep{carusotto13}. However, recently \citet{darmon14} suggest that Rabaud \& Moisy were measuring the angle provided by the peaks of the waves.  By treating the linear problem of flow past a pressure distribution, Darmon \emph{et al.} demonstrate that the apparent wake angle defined by the highest wave peaks is consistent with the experimental observations of Rabaud \& Moisy (including a decrease in angle which scales as the inverse object speed), thus describing the trends in the data without making any assumptions about the relationship between wavelengths and the hull length.  This research is reviewed by \citet{dias14}.

A series of very recent papers has followed the work of \citet{rabaud13} and  \citet{darmon14}.  \citet{ellingsen14} considers the problem of flow past a pressure distribution in a shear flow, \citet{benzaquen14} treat elliptic pressure distributions, while \citet{rabaud14} make connections with wave drag for the same type of flow configuration.  In each of these three cases, apparent wake angles are calculated using exact solutions to linear problems.
In a slightly different vein, \citet{noblesse14} proposes another explanation for the decrease in apparent wake angle by suggesting that smaller angles are due to interference between bow and stern waves, providing an additional regime in which the angle appears to scale like the inverse square of the ship speed.  Finally, \citet{moisy14} go some way to connect the work of \citet{noblesse14} with the other recent papers by focusing on flows past pressure distributions with large aspect ratios, while \citet{he14} assess the key assumptions behind each of  \citet{rabaud13}, \citet{darmon14} and \citet{noblesse14} in the context of Rabaud \& Moisy's original observations of real ships.

All of the above theoretical work is for {\em linear} flows, while much of the focus of the present paper is on the effect that {\em nonlinearity} has on the apparent wake angle. We are going to approach this study by considering two flow configurations, the first being the flow due to a point source submerged below the surface in a fluid moving at constant velocity.  For this geometry, the problem is closely related to flow past a submerged semi-infinite Rankine-type body.  There is no obvious length scale in the direction of travel and thus we completely avoid issues related to wave interference between a ship's bow and stern or the question of disturbances having wavelengths greater than or less than the ship hull length.  The second configuration we consider is flow past a submerged source doublet.  Here there is a correspondence with flow past a submerged sphere, although for nonlinear flows care must be taken with this analogy, since the nonlinear case is not equivalent to flow past a closed body.

The mathematical problems we consider are formulated in Section~\ref{sec:formulation}.  The method we employ to measure the apparent wake angle, similar to that used in \citet{moisy14}, is outlined in Section~\ref{sec:measure}.  It is instructive to begin with the linear version of our problems in Section~\ref{sec:linear}, highlighting the key features for comparison with the nonlinear results.  We describe three different regimes.  The first is for small Froude numbers, for which the divergent waves are not visible and so our method does not apply.  The second regime corresponds to wave patterns that have the maximum peaks on the outermost divergent waves, giving rise to apparent wake angles close the Kelvin angle.  For the third regime, the highest peaks form a V-shape with an angle that decreases with the speed of the steadily moving source, scaling as the inverse Froude number in the large Froude number limit.  We then generate numerical solutions to the nonlinear problems in Section~\ref{sec:nonlinear} and observe that nonlinear wave patterns can appear qualitatively different to their linear analogues.  Our results suggest that, for a fixed Froude number, the apparent wake angle increases as the nonlinearity increases, and in some cases continues to increase {\em past the Kelvin angle}.  Furthermore, for large Froude numbers, strong nonlinearity can have the dramatic effect of shifting the regime to be the second type instead of the third.  We conclude that both the speed (the Froude number) and `size' (the nonlinearity) of a moving ship or submarine influence apparent wake angle. In particular, for sufficiently fast-moving ships, the apparent wake angle tends to decrease with ship speed but increase with ship `size'.

We mention in passing there are many other physical influences not treated here.  For example, the consideration of how surface tension \citep{carusotto13,doyle13,moisy14b} or a finite-depth channel \citep{soomere07} affects the wake angle raises additional questions that are worth exploring.  We leave these issues for further study.

\section{Mathematical formulation}
\label{sec:formulation}

\subsection{Governing equations for flow past a point source}

We first consider the problem of steady flow caused by submerging a source of strength $m$ at a depth $H$ below the surface of a uniform stream of speed $U$ travelling in the $x$-direction.  The main task is to solve for the velocity potential, $\phi(x,y,z)$, and determine the shape of the free surface, $z=\zeta(x,y)$.  There are two dimensionless parameters: the dimensionless source strength, $\epsilon=m/(UH^2)$ and the depth based Froude number, $F=U/\sqrt{gH}$, where here $g$ is acceleration due to gravity.  For a small source strength ($\epsilon\ll 1$), the problem can be linearised about the undisturbed surface ($z=0$) \citep{noblesse78,chung91,lustrichapman13}, while for moderate to large source strengths, the full nonlinear problem must be solved in the domain $z<\zeta(x,y)$ \citep{forbes89,pethiyagoda14}.  We present the dimensionless governing equations for both cases.

Assuming ideal fluid flow conditions, the velocity potential $\phi$ satisfies Laplace's equation below the free-surface:
\begin{equation}
\nabla^2\phi=\frac{\partial^2 \phi}{\partial x^2}+\frac{\partial^2 \phi}{\partial y^2}+\frac{\partial^2 \phi}{\partial z^2}=0 \quad \mbox{for\ }z<\zeta(x,y)\mbox{ (nonlinear) or\ }z<0\mbox{ (linear)}.\label{eqn:NondimLap}
\end{equation}
We have kinematic and dynamic conditions acting on the free-surface:
\begin{align}
\phi_x\zeta_x+\phi_y\zeta_y=\phi_z \mbox{\ on\ } z=\zeta(x,y),\label{eqn:NondimKin}\\
\frac{1}{2}(\phi^2_x+\phi^2_y+\phi^2_z)+\frac{\zeta}{F^2}=\frac{1}{2} \mbox{\ on\ } z=\zeta(x,y),\label{eqn:NondimDyn}
\end{align}
for the nonlinear problem.  The kinematic condition states that no fluid can flow through the surface $z=\zeta(x,y)$, while the dynamic condition combines Bernoulli's equation with a constant atmospheric pressure on $z=\zeta(x,y)$.  After linearisation, these two equations become
\begin{align}
\zeta_x&=\phi_z \quad \text{on} \quad z=0,\label{eqn:NondimKinlinear}\\
\phi_x-1+\frac{\zeta}{F^2}&=0 \quad\;\: \text{on} \quad z=0,\label{eqn:NondimDynlinear}
\end{align}
respectively. The velocity potential near the source behaves like
\begin{equation}
\phi\sim-\frac{\epsilon}{4\pi\sqrt{x^2+y^2+\left(z+1\right)^2}} \mbox{\ as\ } (x,y,z)\rightarrow(0,0,-1).\label{eqn:NondimSource}
\end{equation}
Finally, we have far field conditions both infinitely far upstream and below the surface,
\begin{align}
(\phi_x,\phi_y,\phi_z)\rightarrow(1,0,0),&\quad \zeta \rightarrow 0\quad\text{as} \quad x\rightarrow-\infty,\label{eqn:NondimUp}
\\
(\phi_x,\phi_y,\phi_z)\rightarrow(1,0,0),&\qquad\qquad\;\;\text{as}\quad z\rightarrow-\infty.\label{eqn:NondimFar}
\end{align}
The governing equations (\ref{eqn:NondimLap})--(\ref{eqn:NondimFar}) define both the nonlinear and linear problems for free-surface flow past a submerged source.  Once the solution for $\phi$ is determined, the velocity field ${\bf q}$ is recovered via ${\bf q}=\nabla\phi$.

The linear problem (\ref{eqn:NondimLap}), (\ref{eqn:NondimKinlinear})--(\ref{eqn:NondimFar}) above is equivalent to flow past a submerged semi-infinite Rankine body with a rounded nose \citep[pp. 461]{batchelor}.  As such, one interpretation is that this problem models the ship wave pattern generated at the bow of a submerged body such as a submarine.  Downstream from the nose, the body very quickly approaches a cylinder in shape with cross-sectional radius $\sqrt{\epsilon/\pi}$ and area $\epsilon$. Thus, with our choice of nondimensionalisation, we are able to use the following variation of parameter values to get a sense of the physics involved with this choice of flow configuration.  First, we see that increasing (decreasing) the Froude number $F$ while keeping the dimensionless source strength $\epsilon$ fixed is equivalent to increasing (decreasing) the speed of the flow about the Rankine body while keeping the size and depth of the Rankine body fixed.   On the other hand, increasing (decreasing) $\epsilon$ while keeping $F$ fixed is equivalent to increasing (decreasing) the size of the Rankine body while keeping its depth and the fluid speed fixed.  Finally, if we increase (decrease) both $F$ and $\epsilon$ such that $\epsilon/F^4$ is held constant, then this is equivalent to fixing the speed of the flow and the size of the Rankine body, but decreasing (increasing) the depth of the body. For this linear problem we shall make use of the exact solution
\begin{align}
\zeta(x,y)=&-\frac{\epsilon F^2\,\mathrm{sgn}(x)}{\pi^2}\int_{0}^{\frac{\pi}{2}}\cos\theta
\int_{0}^{\infty}\frac{k \mathrm{e}^{-k|x|}\cos(k y \sin \theta)g(k,\theta)}{F^4 k^2+\cos^2\theta}\,\,\mbox{d}k\,\mbox{d}\theta
\nonumber\\
&+\frac{\epsilon H(x)}{\pi}\int_{-\infty}^{\infty}\xi \mathrm{e}^{-F^2\xi^2}\cos(x\xi)\cos(y\xi \lambda)\,\,\mbox{d}\lambda
\label{eq:sourcefunc}
\end{align}
\citep{peters49},
where
\begin{align}
g(k,\theta)&=F^2 k \sin(k\cos \theta)+\cos\theta\cos(k\cos\theta),\label{eq:g}\\
\xi(\lambda)&=\sqrt{\lambda^2+1}/F^2,
\label{eq:xi}
\end{align}
$\mathrm{sgn}(x)$ is the sign function, and $H(x)$ is the Heaviside function. The integrals in this exact solution are evaluated numerically to obtain the free-surface profile for a given Froude number, $F$.

The nonlinear problem (\ref{eqn:NondimLap})--(\ref{eqn:NondimDyn}), (\ref{eqn:NondimSource})--(\ref{eqn:NondimFar}) is also equivalent to flow past a semi-infinite body, although there is no longer a well-known formula for the body shape.  This cigar-like body will have a cross-sectional area in the far-field which is roughly of order $\epsilon$, at least for moderate values of $\epsilon$.  For large $\epsilon$, the shape of the submerged body will be highly distorted, and the direct correspondence with the semi-infinite Rankine body is lost.

\subsection{Governing equations for flow past a source doublet}

The second problem we are concerned with is free surface flow past a source doublet of strength $\kappa$ submerged a depth $H$ below the surface of a uniform stream of speed $U$ \citep{havelock28,havelock32b}. This problem has two dimensionless parameters, the source doublet strength, $\mu=\kappa/(UH^3)$, and the Froude number, $F=U/\sqrt{gH}$. With this scaling, the governing equations for flow past a doublet are the same for flow past a source, except that (\ref{eqn:NondimSource}) is replaced by
\begin{align}
\phi\sim&\frac{\mu x}{4\pi\left(x^2+y^2+\left(z+1\right)^2\right)^{3/2}} \mbox{\ as\ } (x,y,z)\rightarrow\left(0,0,-1\right).\label{eqn:NondimDipole}
\end{align}

The linearised problem of flow past a doublet (\ref{eqn:NondimLap}), (\ref{eqn:NondimKinlinear})--(\ref{eqn:NondimDynlinear}), (\ref{eqn:NondimUp})--(\ref{eqn:NondimFar}), (\ref{eqn:NondimDipole}) is valid for the weak doublet limit $\mu\ll 1$. In this case the problem is equivalent to flow past a submerged sphere of radius $(\mu/2\pi)^{1/3}$ and volume $2\mu/3$ \citep[pp. 116]{lamb16}. Thus, increasing the Froude number $F$ while keeping $\mu$ constant is equivalent to increasing the velocity in the far field while keeping the radius and depth of the submerged sphere fixed.  Increasing $\mu$ with $F$ fixed is like increasing the radius of the sphere while holding its depth and the far-field speed constant.  Finally, increasing both $F$ and $\mu$ with $\mu/F^6$ constant is equivalent to fixing the speed of the flow and radius of the sphere while decreasing the sphere's depth.  An exact solution this linear problem is given by
\begin{align}
\zeta(x,y)=&\frac{\mu F^2}{\pi^2}\int_{0}^{\frac{\pi}{2}}\cos\theta
\int_{0}^{\infty}\frac{k^2 \mathrm{e}^{-k|x|}\cos(k y \sin \theta)g(k,\theta)}{F^4 k^2+\cos^2\theta}\,\,\mbox{d}k\,\mbox{d}\theta
\nonumber\\
&-\frac{\mu H(x)}{\pi}\int_{-\infty}^{\infty}\xi^2 \mathrm{e}^{-F^2\xi^2}\sin(x\xi)\cos(y\xi \lambda)\,\,\mbox{d}\lambda,
\label{eq:doubletfunc}
\end{align}
where $g(k,\theta)$ and $\xi(\lambda)$ are given by (\ref{eq:g}) and (\ref{eq:xi}), respectively. As with flow past a source, nonlinearity has the effect of distorting the shape of the submerged object.  In fact, for nonlinear flow past a doublet, the problem is not equivalent to flow past a closed object, as there are no closed streamsurfaces.  Regardless, this nonlinear problem is still of considerable interest in the present context, as there is no net increase of fluid in the system, which contrasts the problem of flow past a point source.  Further, the nonlinear wave patterns and resulting wake angles are likely to be closely related to those associated with flow past submerged closed bodies  \citep[see][for example]{tuck02}.


\section{Method of measuring the apparent wake angle}\label{sec:measure}

\begin{figure}
\centering
\includegraphics[width=.55\linewidth]{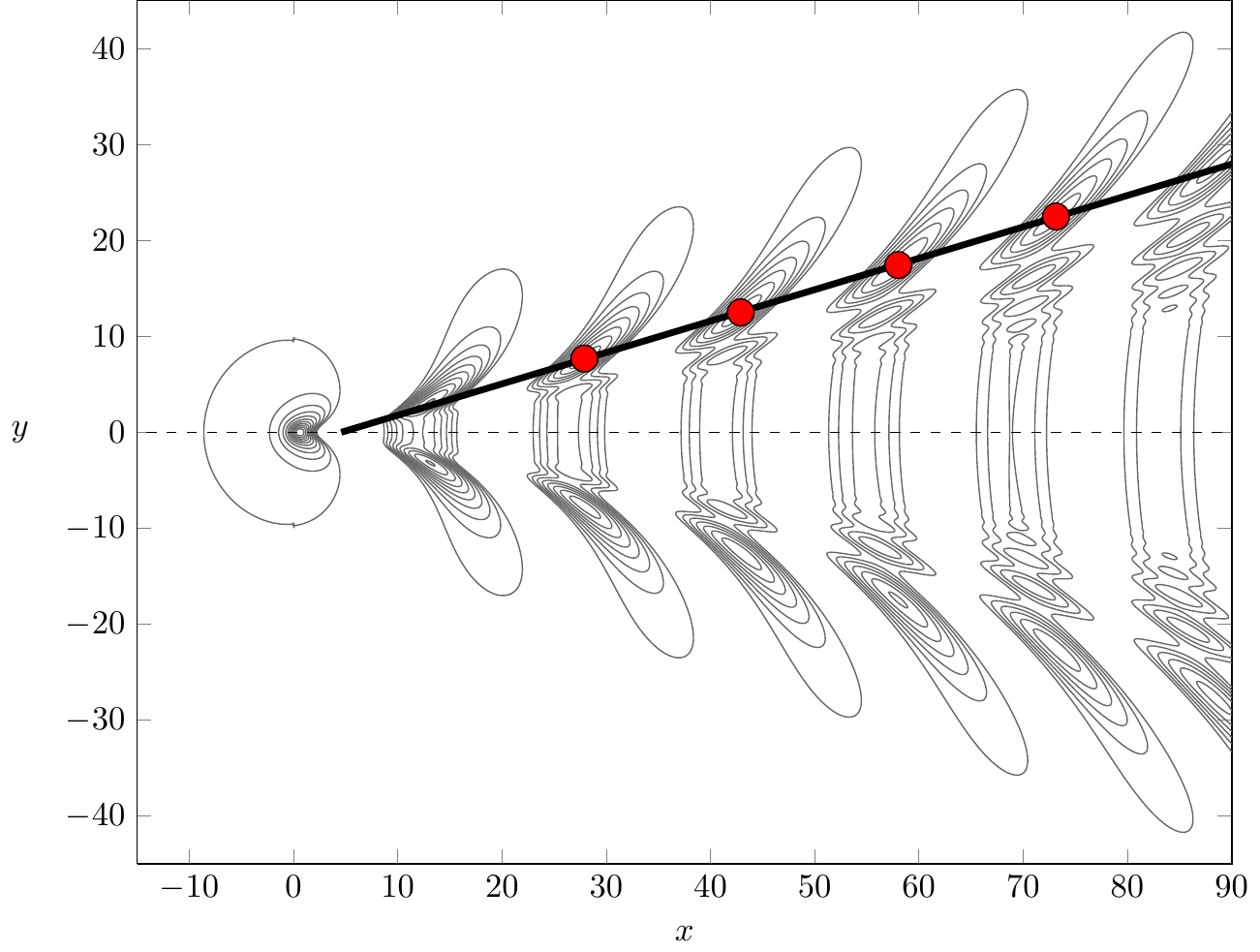}
\caption{Contour plot of the linear free-surface profile for flow past a source with $F=1.5$, with only positive wave heights shown for clarity. The highest wake peaks are marked with solid (red) circles. The thick (black) line represents the line of best fit.}
\label{fig:ContourPeakFr15full}
\end{figure}

In order to measure the apparent wake angle, $\theta_\mathrm{app}$, of a given free-surface profile, we isolate every wavelength of the wave and mark the highest peak. We then fit a line through these points with a least-squares approximation. Finally, we take the angle between the line of best fit and the centreline ($y=0$) as the apparent wake angle. An illustration of this method is shown in figure~\ref{fig:ContourPeakFr15full} for linearised flow past a point source with $F=1.5$. This method is also used very recently by \citet{moisy14}; it acts to approximate the manual approach of \citet{rabaud13}, who drew a line through the brightest points in a series of satellite images of ship wakes and measured the resulting angle by hand.

On the other hand, an alternative approach for measuring the apparent wake angle is employed by \citet{darmon14} and \citet{benzaquen14}, who define their wake angle in terms of the maximum of a curve which is interpolated from the peaks within a given wavelength far downstream.  A line is drawn between this maximum and the origin (which, for their problem, is the centre of an applied pressure distribution), forming an angle with the centreline.  A similar method was used by \citet{ellingsen14}.  While this approach works well for measuring wake angles for linear problems with exact solutions, we are unable to compute nonlinear free-surface profiles far downstream without loss of accuracy.  Thus we prefer our method, which is more straightforward, and involves fitting a line through data points closer to the disturbance in question.



\section{Linear ship waves}\label{sec:linear}
Using the exact solutions, (\ref{eq:sourcefunc}) and (\ref{eq:doubletfunc}) to the linear problems of flow past a source and doublet, respectively, we calculate free-surface profiles for a number of different Froude numbers. By applying our method for measuring the apparent wake angle $\theta_\mathrm{app}$, we have recorded the dependence of $\theta_\mathrm{app}$ on the Froude number $F$ in figure~\ref{fig:FrvsPeak}, which we now discuss in some detail.

\subsection{Measured wake angle for flow past a source}
\begin{figure}
\centering
\subfloat[Source]{\includegraphics[width=.7\linewidth]{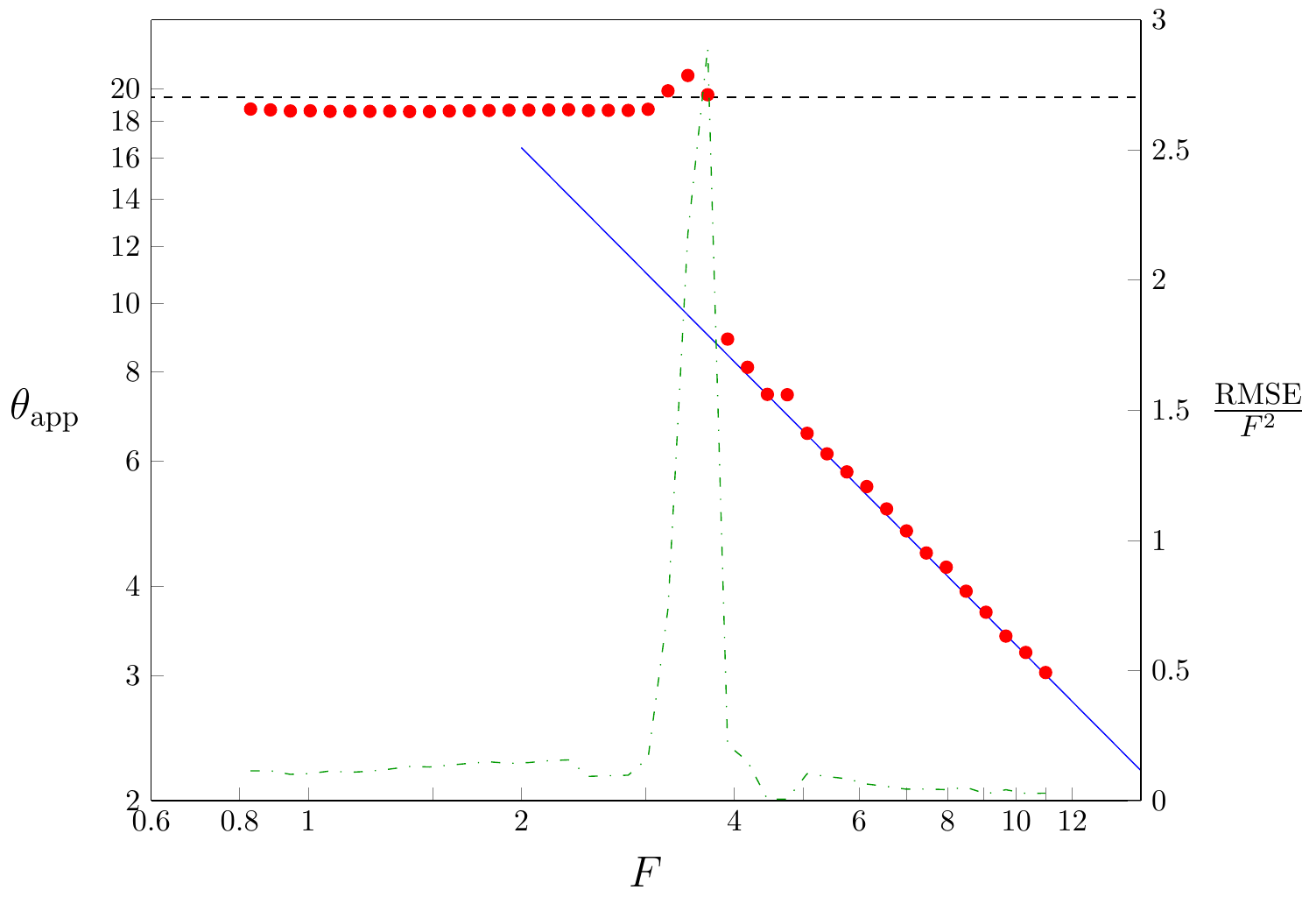}\label{fig:FrvsPeaksource}}\\
\subfloat[Doublet]{\includegraphics[width=.7\linewidth]{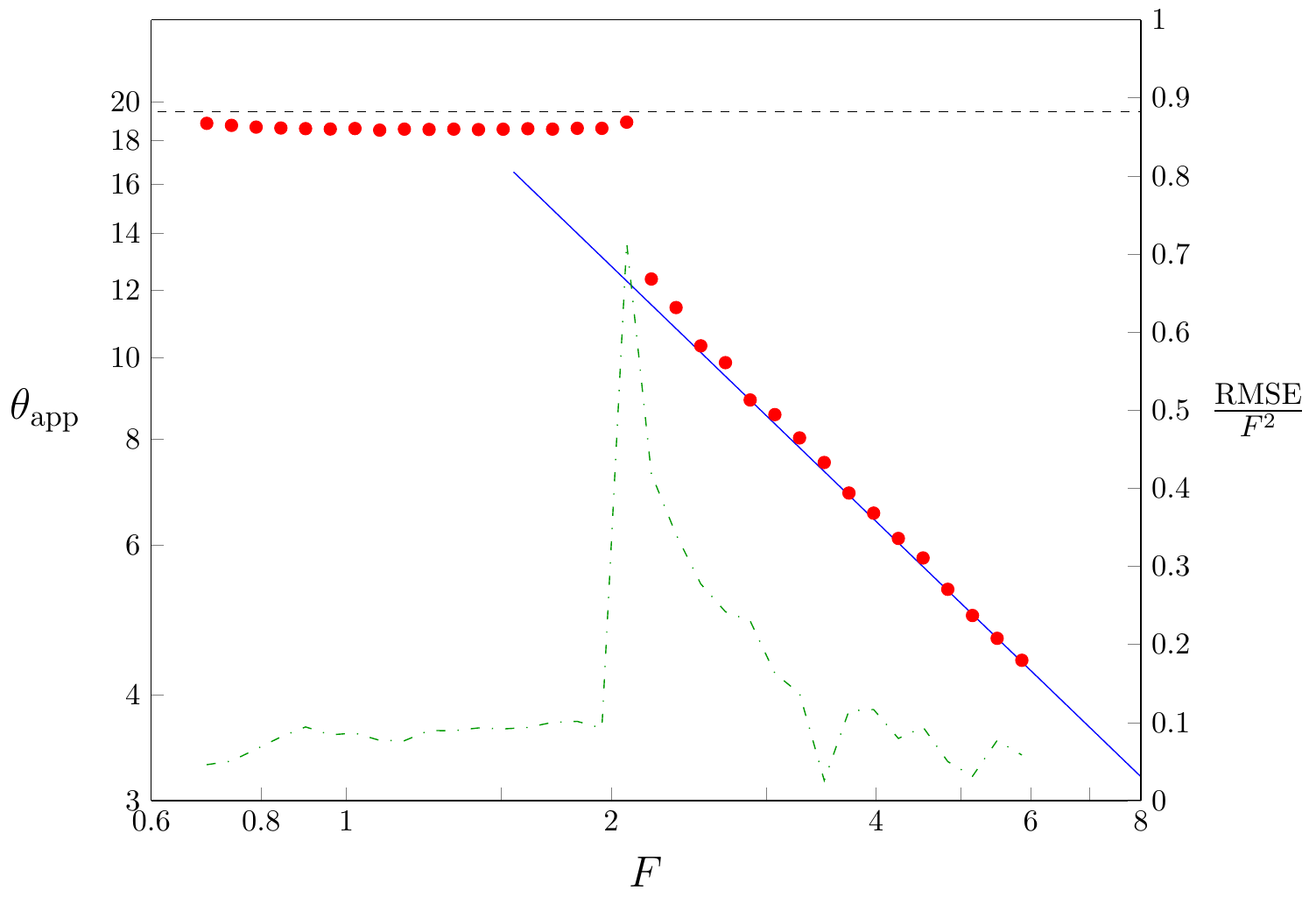}\label{fig:FrvsPeakdoublet}}\\
\caption{Measured apparent wake angles $\theta_\mathrm{app}$ in degrees (solid circles) plotted against the Froude number for the linear solution (\ref{eq:sourcefunc}) and (\ref{eq:doubletfunc}) for flow past (a) a source and (b) a doublet, respectively, presented on a log-log scale. The dashed line is Kelvin's angle $\arcsin(1/3)\approx 19.47^\circ$.  The solid (blue) line is the theoretical asymptote shown in (a) equation (\ref{eq:AngleAsymp}) or (b) equation (\ref{eq:AngleAsympDoublet}), valid for large Froude numbers.  The (green) dot-dashed curve represents the root mean squared error, scaled by $F^2$.}
\label{fig:FrvsPeak}
\end{figure}
\begin{figure}
\centering
\subfloat[$F=1.5$]{\includegraphics[width=.45\linewidth]{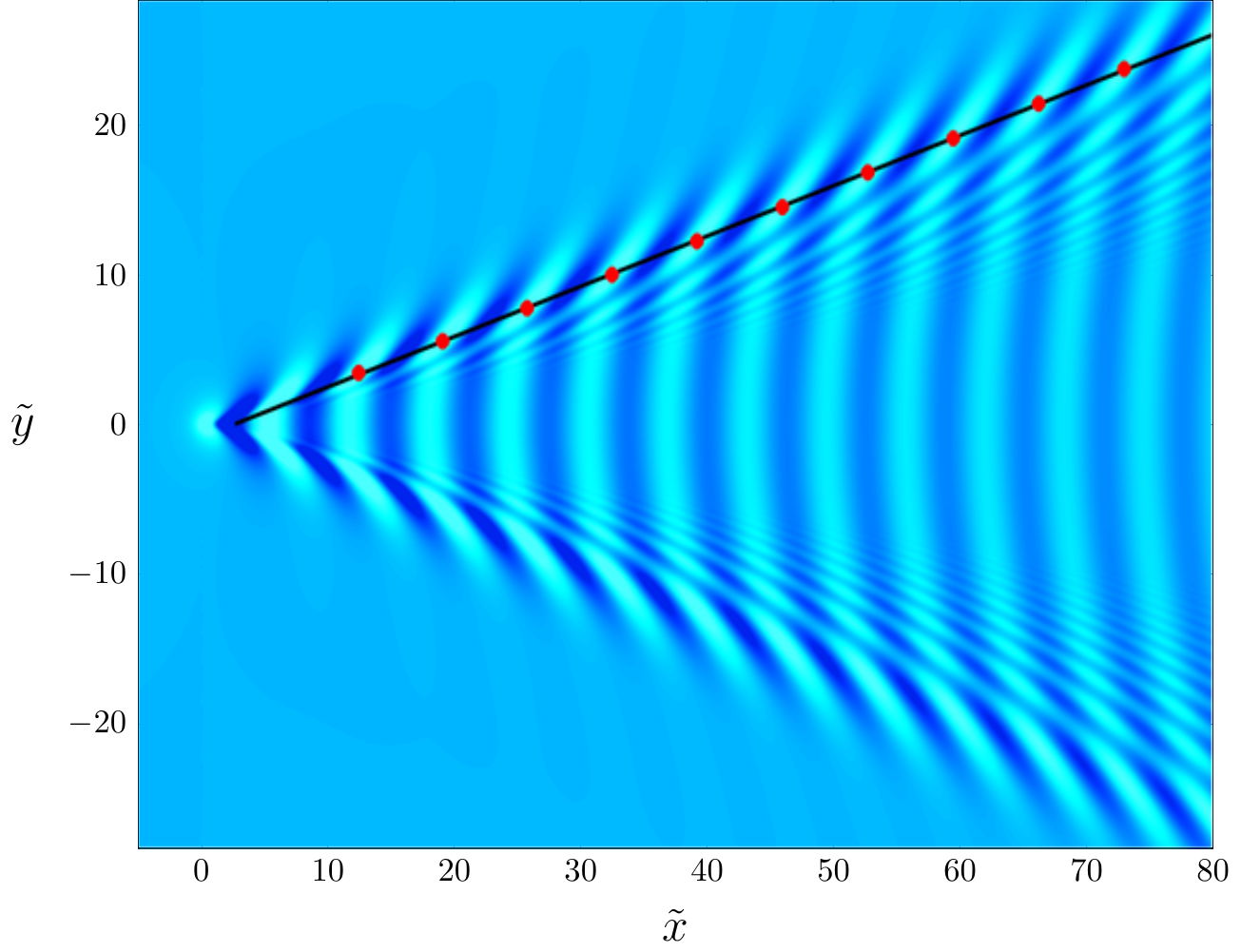}\label{fig:LinSurfFr15}}
\subfloat[$F=3.5$]{\includegraphics[width=.45\linewidth]{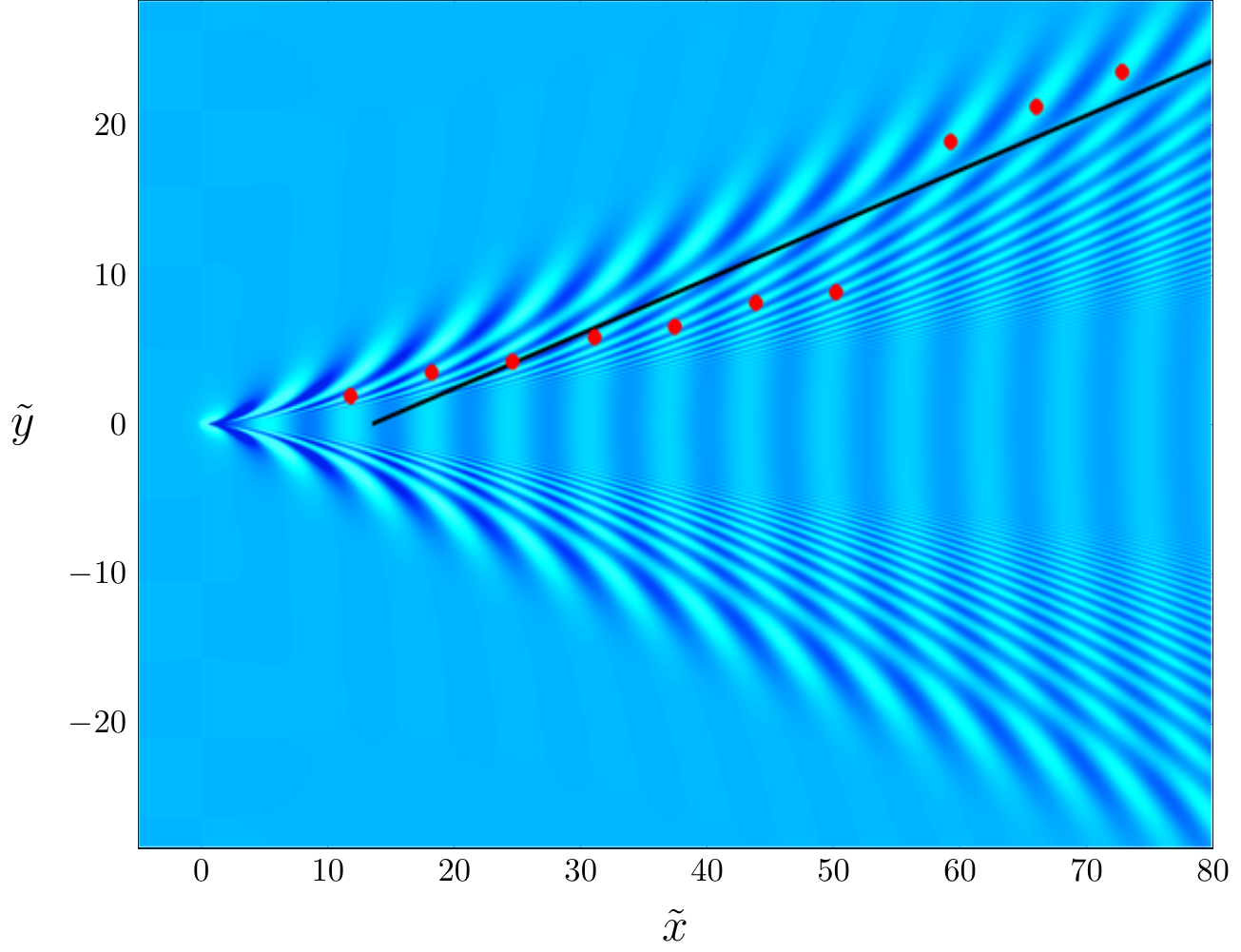}\label{fig:LinSurfFr35}}\\
\subfloat[$F=4.5$]{\includegraphics[width=.45\linewidth]{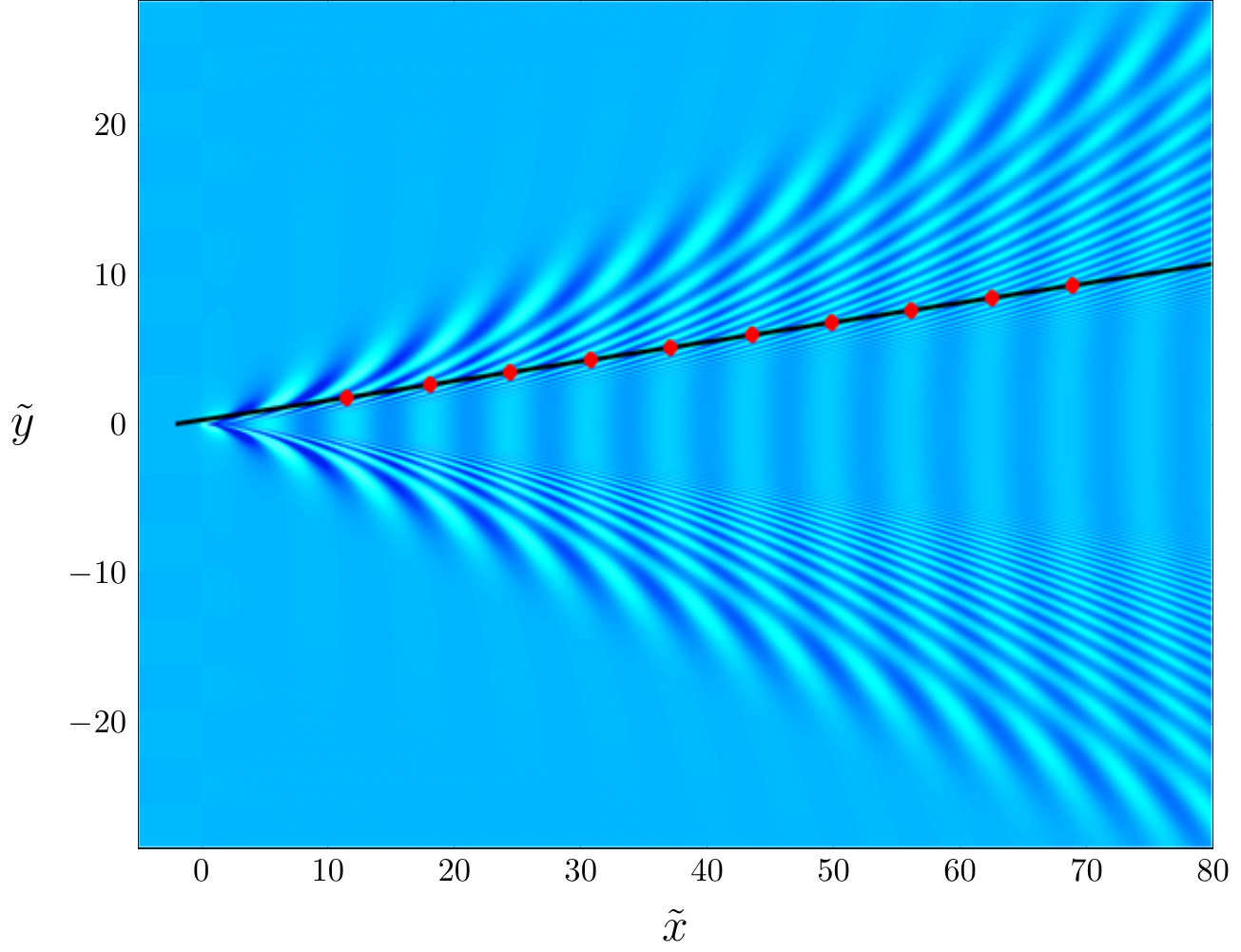}\label{fig:LinSurfFr45}}
\subfloat[$F=8.5$]{\includegraphics[width=.45\linewidth]{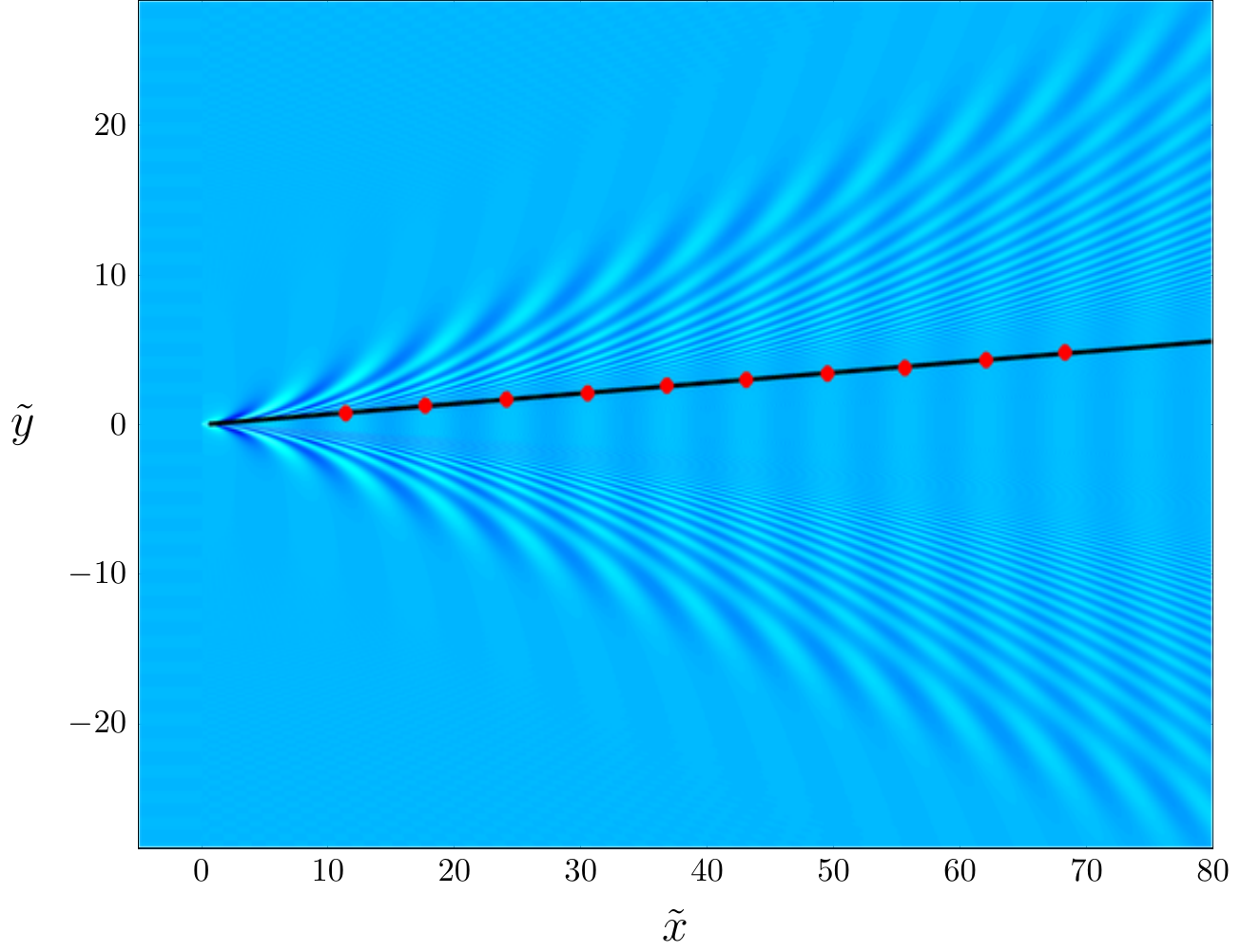}\label{fig:LinSurfFr85}}
\caption{Plan view of the surface elevation in the linear case, computed for four different Froude numbers.
The (red) solid circles are the highest peaks and the solid (black) line is the line of best fit, as described in Section~\ref{sec:measure}. The surfaces are plotted on the scaled domain $\tilde{x}=x/F^2$ and $\tilde{y}=y/F^2$, for ease of comparison.}\label{fig:LinSurf}
\end{figure}

Figure~\ref{fig:FrvsPeak}(a) shows results for flow past a source. For the approximate range of Froude numbers $0.8<F<3$, we see the apparent wake angle remains roughly constant at about $18.5^\circ$, which is slightly less than the Kelvin angle.  We call this Regime II, and show in figure~\ref{fig:LinSurf}(a) a typical wave pattern from this parameter range.  As with other solutions in Regime II, the wave pattern in figure~\ref{fig:LinSurf}(a) is characterised by the highest peaks lying on the outermost divergent waves.

For sufficiently large Froude numbers, we find that the highest peaks on the wave pattern no longer lie on the outermost divergent waves, and so the apparent wake angle is much less than Kelvin's angle.  We call this Regime III.  Figures~\ref{fig:LinSurf}(c),(d), which are drawn for $F=4.5$ and 8.5, respectively, provide examples of wave patterns which lie in this regime.  In fact, for $F>4$ we find the apparent wake angle scales as $F^{-1}$, demonstrated by the asymptotic line in figure~\ref{fig:FrvsPeak}(a) (see Section~\ref{subsec:LargeFroude}).  This result is consistent with the scalings discussed in \citet{darmon14}, \citet{ellingsen14}, \citet{benzaquen14} and \citet{moisy14}.
The wave pattern in figure~\ref{fig:LinSurf}(b) falls somewhere between Regimes II and III, since there is small range of Froude numbers for which all peaks downstream lie on the outermost divergent waves, while all peaks upstream do not.  By noting the location of the solid circles in this special borderline case, we see that our approach for approximating the apparent angle gives a slightly higher value together with a large root mean squared error (indicated by the dot-dashed curve). The spike in error in figure~\ref{fig:FrvsPeak}(a) is clearly apparent in this transitional region.

Finally, we note that for sufficiently small Froude numbers, the peaks of the divergent waves are lower than the transverse waves, and in fact for roughly $F<0.6$ the divergent waves are no longer visible for the spatial resolution employed.  As such, our method of measuring the apparent wake angle breaks down.  We call this Regime I.  Roughly speaking, Regime I (and the transition between Regimes I and II) occurs for $F<0.8$ for this problem.  We shall not consider this regime in the present study, but note it is analysed recently in some detail for the linearised problem of flow past a submerged source in \citet{lustrichapman13}.  The techniques required for this type of small Froude number limit are based on the theory of exponential asymptotics outlined in \citet{chapman06} \citep[and developed further by][for example]{trinh11,lustri12}.

\subsection{Measured wake angle for flow past a doublet}
The relationship between measured apparent wake and the Froude number for flow past a doublet is shown in figure~\ref{fig:FrvsPeak}(b). The plot exhibits the same trends as for flow past a source; the only differences are the intervals for which the regimes are defined. Regime II now appears to hold for values of the Froude number roughly $0.7<F<2$ and Regime III for values $F>2.7$. The transition between these two regimes still appears, but it is not as obvious from figure~\ref{fig:FrvsPeak}(b). While not shown, Regime I occurs for roughly $F<0.5$.

\subsection{Large Froude number approximation}
\label{subsec:LargeFroude}

In the exact solution (\ref{eq:sourcefunc}) to the linearised problem of flow past a source, the second integral dominates the first far downstream.  As such, we can perform a stationary phase approximation on the second integral to provide an analytical approximation to the wave pattern.  Thus for large $r=\sqrt{x^2+y^2}$ we obtain the far field approximation in polar coordinates:
\begin{equation}
\zeta(r,\theta) \sim a_1(r,\theta)\cos\left(r g(\lambda_1(\theta),\theta)+\frac{\pi}{4}\right)
+a_2(r,\theta)\cos\left(r g(\lambda_2(\theta),\theta)-\frac{\pi}{4}\right)
\nonumber
\label{eq:sphase}
\end{equation}
as $r\rightarrow\infty$ for $|\theta|<\arcsin(1/3)$, where
\begin{align}
\lambda_1(\theta)&=\frac{-1+\sqrt{1-8\tan^2\theta}}{4\tan\theta},
&\lambda_2(\theta)&=\frac{-1-\sqrt{1-8\tan^2\theta}}{4\tan\theta},
\label{eqn:parts1}\\
f(\lambda)&=\frac{\sqrt{\lambda^2+1}}{F^2}\mathrm{e}^{-\frac{\lambda^2+1}{F^2}},
&g(\lambda,\theta)&=\frac{\sqrt{\lambda^2+1}}{F^2}
(\cos\theta+\lambda\sin\theta),
\label{eqn:parts2}\\
a_1(r,\theta)&=\sqrt{\frac{2}{\pi}}\frac{\epsilon f(\lambda_1(\theta))}{\sqrt{r|g_{\lambda\lambda}
(\lambda_1(\theta),\theta)|}},
&a_2(r,\theta)&=\sqrt{\frac{2}{\pi}}\frac{\epsilon f(\lambda_2(\theta))}{\sqrt{r|g_{\lambda\lambda}
(\lambda_2(\theta),\theta)|}},
\label{eqn:parts3}
\end{align}
and $g_{\lambda\lambda}(\lambda,\theta)$ is the second partial derivative of $g(\lambda,\theta)$ with respect to $\lambda$. Here $a_1$ and $a_2$ are the function envelopes of the transverse and divergent waves, respectively.

As the Froude number increases, the divergent waves dominate the wake pattern, thus for large Froude numbers we need only consider the divergent wave ($a_2$) when calculating the apparent wake angle.  Following \citet{darmon14}, we apply a small $\theta$ approximation to (\ref{eqn:parts1})--(\ref{eqn:parts3}) to give
$$
a_2(r,\theta)\sim \frac{\epsilon}{2F\sqrt{\pi r}}
\left(
\theta^{-3/2}+\frac{7}{4}\theta^{1/2}+\ldots\right)
\mathrm{e}^{(\theta^{-2}+2/3+\ldots)/4F^2}
$$
for $\theta\ll 1$.  Differentiating with respect to $\theta$ and setting the result to zero, we find the maximum of $a_2$ satisfies $1+(7/4-3F^2)\theta^2+\ldots=0$, which leads to the leading order expression
\begin{equation}
\theta_\mathrm{app}\sim\frac{1}{\sqrt{3}F}\quad \mbox{as } F\rightarrow\infty.\label{eq:AngleAsymp}
\end{equation}
This result is plotted in figure~\ref{fig:FrvsPeak}(a). We see that the data for $\theta_\mathrm{app}$ agrees very well with this approximation in Regime III.

Similarly, a far field approximation and subsequent analysis of the divergent wave can be made for flow past a doublet using equation (\ref{eq:doubletfunc}). This gives the apparent wake angle for the doublet
\begin{equation}
\theta_\mathrm{app}\sim\frac{1}{\sqrt{5}F}\quad \mbox{as } F\rightarrow\infty.\label{eq:AngleAsympDoublet}
\end{equation}
Again, we have included the asymptotic result in figure \ref{fig:FrvsPeak}(b) and the agreement with the data in Regime III is very good. This scaling $\theta_\mathrm{app}=O(F^{-1})$ as $F\rightarrow\infty$ for flow past submerged bodies is the same as that found for flows past pressure distributions on the surface \citep{benzaquen14,darmon14,ellingsen14,moisy14}.
\section{Nonlinear ship waves}\label{sec:nonlinear}

The fully nonlinear problems for flow past a source (\ref{eqn:NondimLap})--(\ref{eqn:NondimDyn}), (\ref{eqn:NondimSource})--(\ref{eqn:NondimFar}) and flow past a doublet (\ref{eqn:NondimLap})--(\ref{eqn:NondimDyn}), (\ref{eqn:NondimUp})--(\ref{eqn:NondimFar}), (\ref{eqn:NondimDipole}) do not have known exact solutions, and must be solved numerically.  The approach used here is based on a boundary integral method developed by \citet{forbes89} and
\citet{parau02} which involves setting $\Phi(x,y)=\phi(x,y,\zeta(x,y))$ and formulating the integral equation
\begin{align}
2\pi(\Phi(x,y)-x)=& \, S(x,y) +\int\limits_{0}^{\infty}\int\limits_{-\infty}^{\infty}(\Phi(x^*,y^*)-\Phi(x,y)-x^*+x)K_1(x^*,y^*;x,y)\,\, \text{d}x^*\,\text{d}y^*\notag\\
&+\int\limits_{0}^{\infty}\int\limits_{-\infty}^{\infty}\zeta_x(x^*,y^*)K_2(x^*,y^*;x,y)\,\, \text{d}x^*\,\text{d}y^*,\label{eqn:IntegroEqn}
\end{align}
where $K_1$ and $K_2$ are the kernel functions
\begin{align*}
K_1(x^*,y^*;x,y)=&\frac{\zeta(x^*,y^*)-\zeta(x,y)-(x^*-x)\zeta_x(x^*,y^*)-(y^*-y)\zeta_y(x^*,y^*)}
{\Bigl((x^*-x)^2+(y^*-y)^2+\bigl(\zeta(x^*,y^*)-\zeta(x,y)\bigr)^2\Bigr)^\frac{3}{2}}\\
&+\frac{\zeta(x^*,y^*)-\zeta(x,y)-(x^*-x)\zeta_x(x^*,y^*)-(y^*+y)\zeta_y(x^*,y^*)}
{\Bigl((x^*-x)^2+(y^*-y)^2+\bigl(\zeta(x^*,y^*)-\zeta(x,y)\bigr)^2\Bigr)^\frac{3}{2}},\\
K_2(x^*,y^*;x,y)=&\frac{1}{\sqrt{(x^*-x)^2+(y^*-y)^2+\bigl(\zeta(x^*,y^*)-\zeta(x,y)\bigr)^2}}\\ &+\frac{1}{\sqrt{(x^*-x)^2+(y^*+y)^2+\bigl(\zeta(x^*,y^*)-\zeta(x,y)\bigr)^2}},
\end{align*}
and the singular term  is given by
$$
S(x,y)=\left\{
\begin{array}{ll}
\displaystyle -\frac{\epsilon}{\left({x}^2+{y}^2+(\zeta(x,y)+1)^2 \right)^\frac{1}{2}}, & \text{flow past source} \\
\\
\displaystyle\frac{\mu x}{\left({x}^2+{y}^2+(\zeta(x,y)+1)^2 \right)^\frac{3}{2}}, & \text{flow past doublet.}
\end{array}
\right.
$$
We truncate the domain in the $(x,y)$-plane and discretise using a rectangular grid.  Then, by enforcing Bernoulli's equation (\ref{eqn:NondimDyn}) and the integral equation (\ref{eqn:IntegroEqn}) at each of the half-mesh points, we derive a large system of nonlinear algebraic equations \citep[see also][and the references therein for commentary on this approach]{forbes05,parau07b}.  By applying a Jacobian-free Newton-Krylov method to this system, we have been able to significantly refine the mesh used when compared to standard contemporary approaches \citep[see][]{pethiyagoda14}.  While full details are included in that paper, we note that a typical mesh employed here was made up of $721$ grid points in the $x$-direction, and $241$ in the $y$-direction.  Solutions computed on a workstation with a GPU\footnote{2x Intel Xeon E5-2670 CPUs with 2.66 GHz processor, M2090 Nvidia Tesla GPU and 128GB of system memory.} typically took 1-2 hours to converge.

\begin{figure}
\centering
\subfloat[Source, $\epsilon=1.5$]{\includegraphics[width=.9\linewidth]{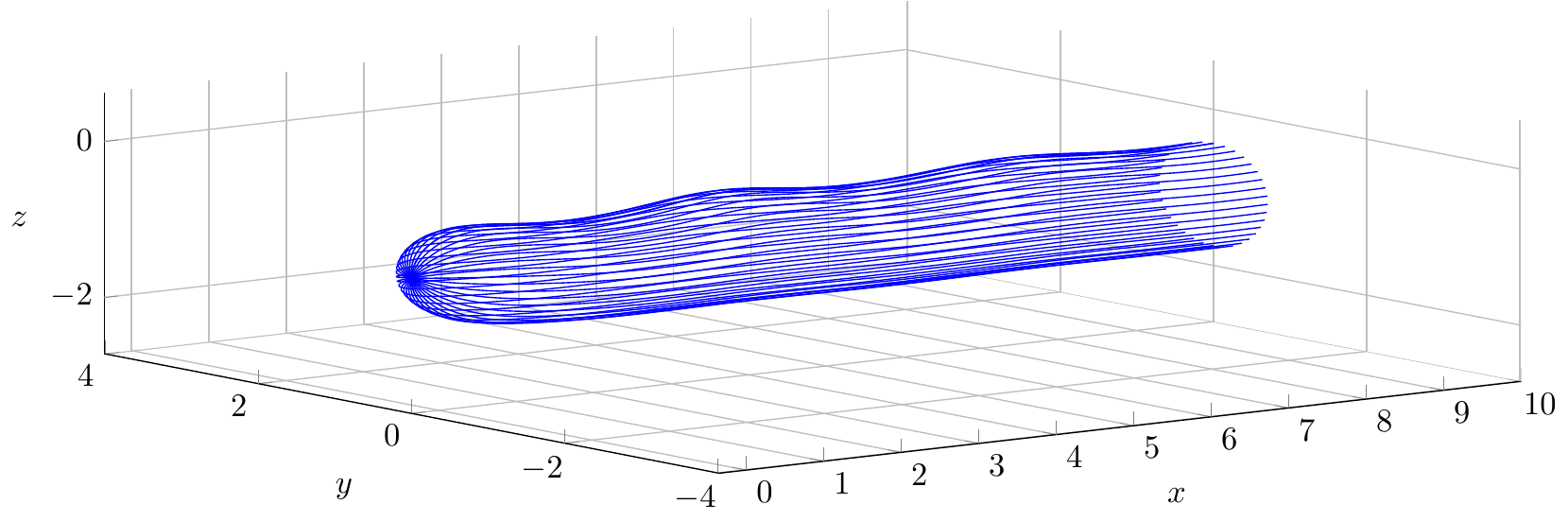}\label{fig:SourceStreamEp15}} \\
\subfloat[Doublet,
$\mu=0.02$]{\includegraphics[width=.5\linewidth]{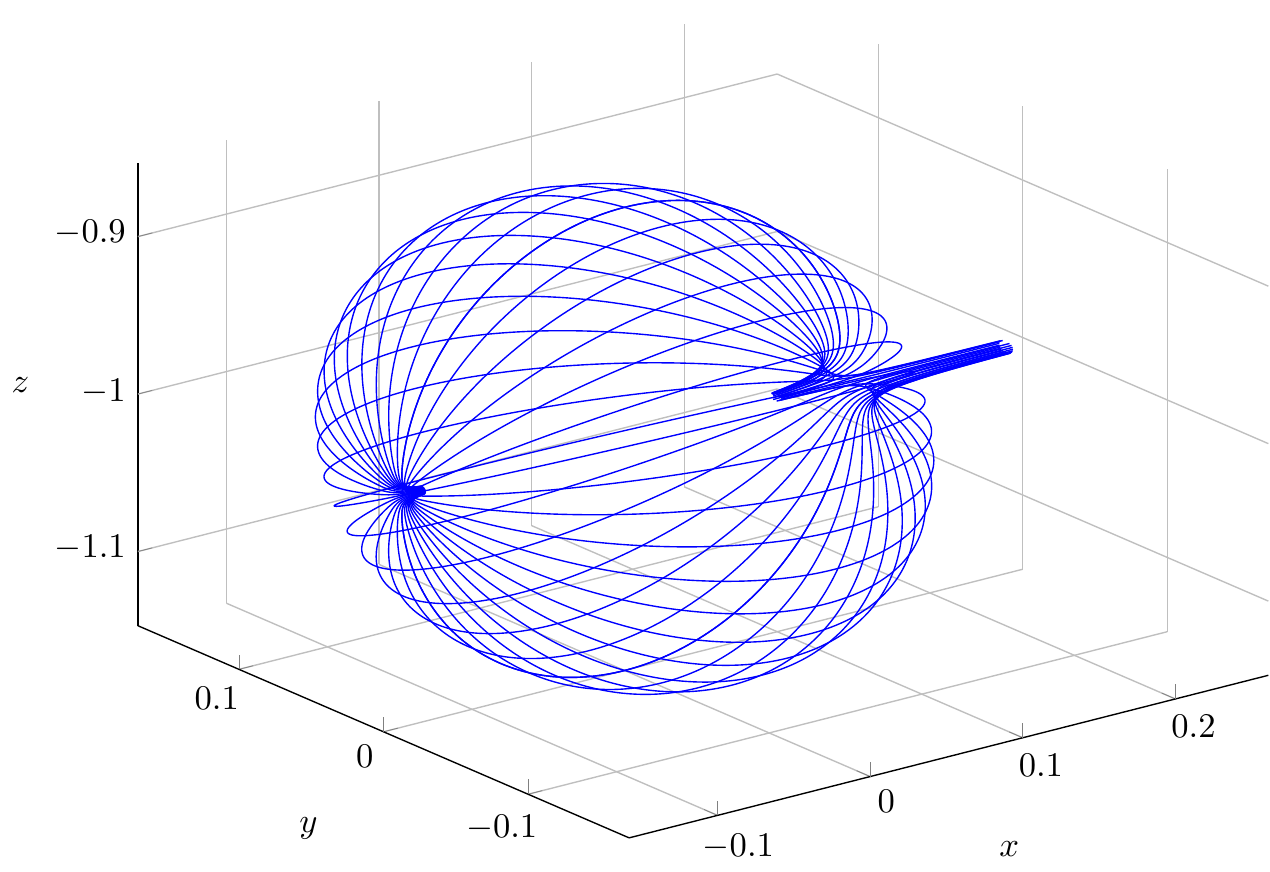}\label{fig:DipoleStream006}}
\subfloat[Doublet, $\mu=0.88$]{\includegraphics[width=.5\linewidth]{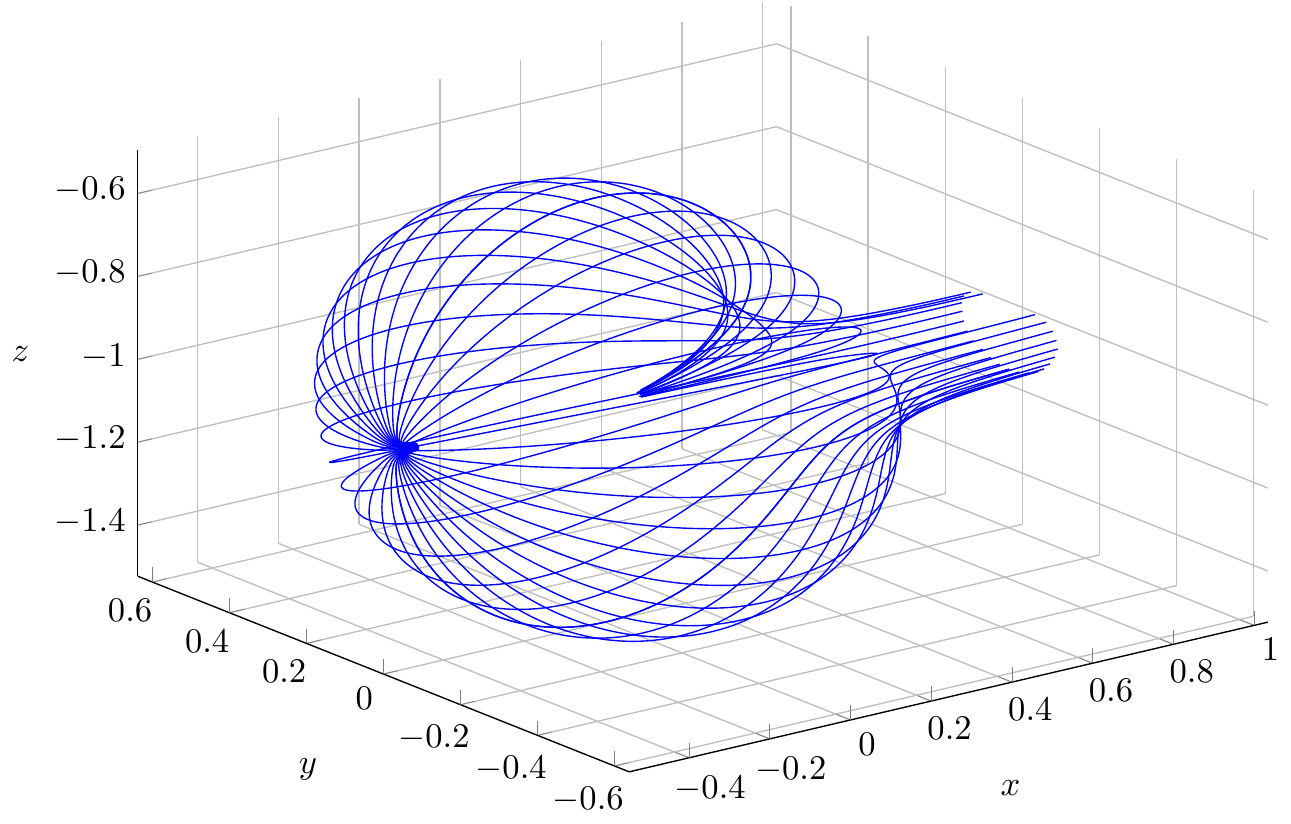}\label{fig:DipoleStreamEp264}}
\\
\subfloat[Doublet,
$\mu=0.02$]{\includegraphics[width=.5\linewidth]{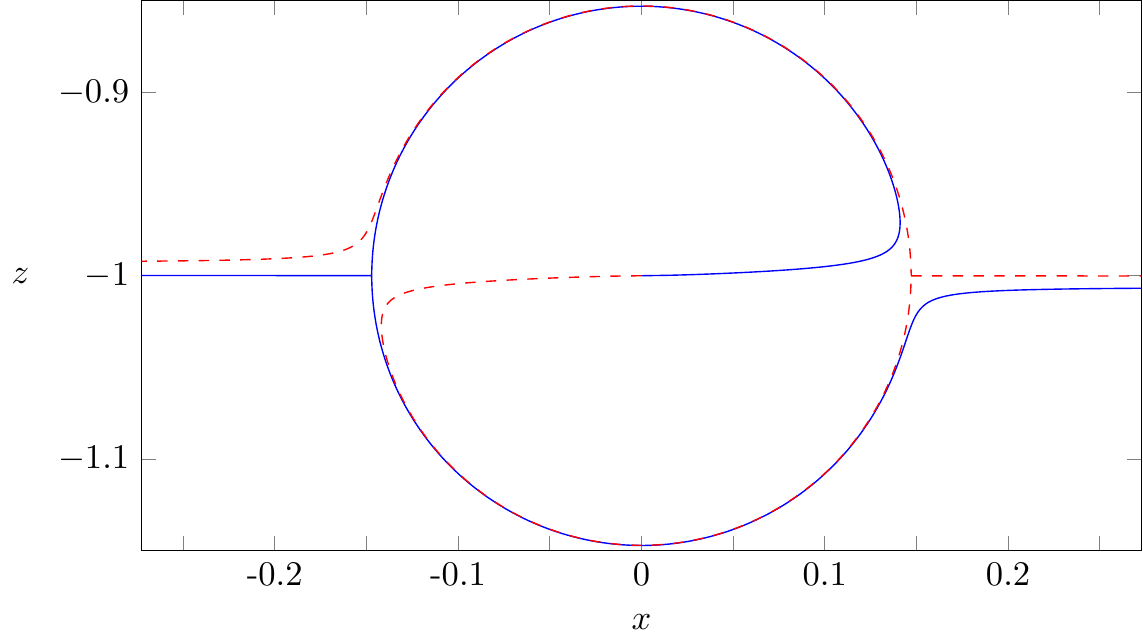}\label{fig:DipoleStream006}}
\subfloat[Doublet, $\mu=0.88$]{\includegraphics[width=.5\linewidth]{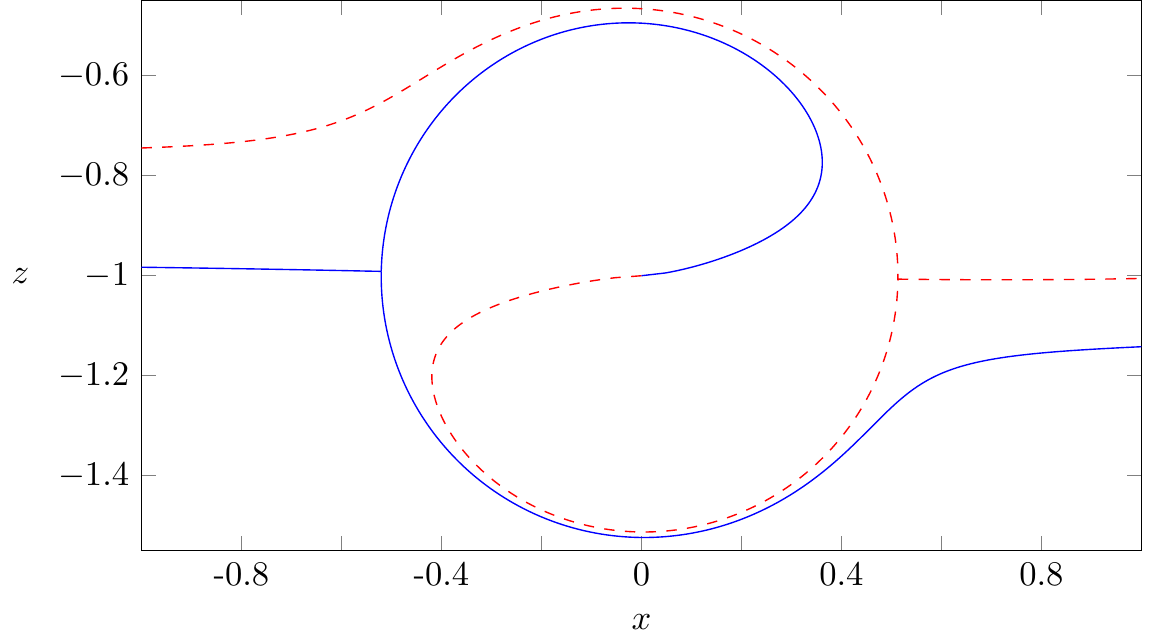}\label{fig:DipoleStreamEp264}}
\\
\caption{Particle paths for flow past a source with (a) $\epsilon=1.5$ and flow past a doublet with (b), (d) $\mu=0.02$ and (c),(e) $\mu=0.88$.  In plots (a)--(c), the particle paths are represented by solid (blue) lines. The particle paths in (d) and (e) lie in the $(x,z)$-plane, with the (blue) solid lines passing through the  upstream stagnation point and the (red) dashed lines passing through the downstream stagnation point.
Each nonlinear solution computed for $F=0.8$.}\label{fig:DipoleStream}
\end{figure}

Recall there are two parameters in each of our problems, one of which is the Froude number $F$. For flow past a source the other parameter is $\epsilon$, while for flow past a doublet it is $\mu$.  Linear solutions are valid for $\epsilon\ll 1$ or $\mu\ll 1$, while for moderate and large values of $\epsilon$ or $\mu$, the full nonlinear problem must be treated.  As discussed in Section~\ref{sec:formulation}, fully nonlinear solutions to these problem no longer have an exact correspondence with their classical linear interpretations.  For example, in Figure~\ref{fig:DipoleStream}(a) we show some particle paths for a nonlinear solution for flow past a submerged source with source strength $\epsilon=1.5$.  These are computed using the boundary integral formulation, which allows the velocity potential $\phi$ to be evaluated at each point within the flow domain (using an equation that is very similar to (\ref{eqn:IntegroEqn}), but has the $2\pi$ on the left-hand side replaced by $4\pi$, and is evaluated within the flow field, not on the surface).  Given a starting point and the velocity potential $\phi$, it is straight forward to calculate the velocity via finite differences and then determine the particle paths using a modified version of the adaptive pathline-based particle tracking algorithm detailed in \citet{bensabat00}.  In Figure~\ref{fig:DipoleStream}(a), a number of starting points is chosen to be very close to the stagnation point within the flow field.  The resulting family of particle paths provide an approximate picture of a submerged body over which there is a no-flux condition, meaning that for these parameters, the problem of flow past a point source is equivalent to flow past this submerged semi-infinite body.  It is only in the linear limit $\epsilon\ll 1$ that this surface becomes an exact semi-infinite Rankine body with radius $\sqrt{\epsilon/\pi}$.

Similarly, figures~\ref{fig:DipoleStream}(b) and (c) show particle paths for flow past a submerged doublet of strength $\mu=0.02$ and $0.88$, respectively. Again, in each case a collection of starting points is chosen very close the upstream stagnation point in the fluid (for this problem there are two stagnation points within the flow domain).  Unlike for the linear solution valid for $\mu\ll 1$ (for which the solutions are equivalent to flow past a submerged sphere of radius $(\mu/2\pi)^{1/3}$), these particle paths do not follow the surface of a closed body.  A complementary picture is provided by figures~\ref{fig:DipoleStream}(d) and (e), which show particle paths in the $(x,z)$-plane.  In this case we show only particle paths that intersect the two stagnation points. None of these curves in this centre-plane join the two stagnation points, again demonstrating the flow field is not identical to that due to a closed submerged body \citep[see][for a discussion on the two-dimensional analogue]{tuck65}.

It is worth clarifying here that, for a fixed Froude number $F$, there is an upper limit on the parameters $\epsilon$ and $\mu$, which we label $\epsilon_{c}$ and $\mu_{c}$, respectively.  For strongly nonlinear flows, the maximum peak height $\zeta_{\mathrm{max}}$ will increase, until $\zeta_{\mathrm{max}}\rightarrow F^2/2$ as either $\epsilon\rightarrow \epsilon_{c}$ (flow past a source) or $\mu\rightarrow \mu_{c}$ (flow past a doublet).  The solution for either $\epsilon= \epsilon_{c}$ or $\mu=\mu_{c}$ will involve some kind of complicated limiting surface configuration, which will include a stagnation point at the highest peak, and possibly even a discontinuous first derivative there (the dynamic condition (\ref{eqn:NondimDyn}) excludes the possibility of solutions that have $\zeta_{\mathrm{max}}> F^2/2$, as the square of the velocity can never be negative).  The two-dimensional analogue is well-studied, with the Stokes limiting configuration characterised by a $120^\circ$ angle at the wave crest; in that case, highly accurate calculations of near-limiting waves are made possible because one is able to treat a single wavelength with periodic boundary conditions \citep{schwartz74,cokelet77,williams81,Lukomsky02,dallaston10}.  On the other hand, the limiting configuration for a given fully three-dimensional flow problem is still relatively unexplored.  We do not pursue this phenomenon here.

\begin{figure}
\centering
\subfloat[Linear]{\includegraphics[width=.45\linewidth]{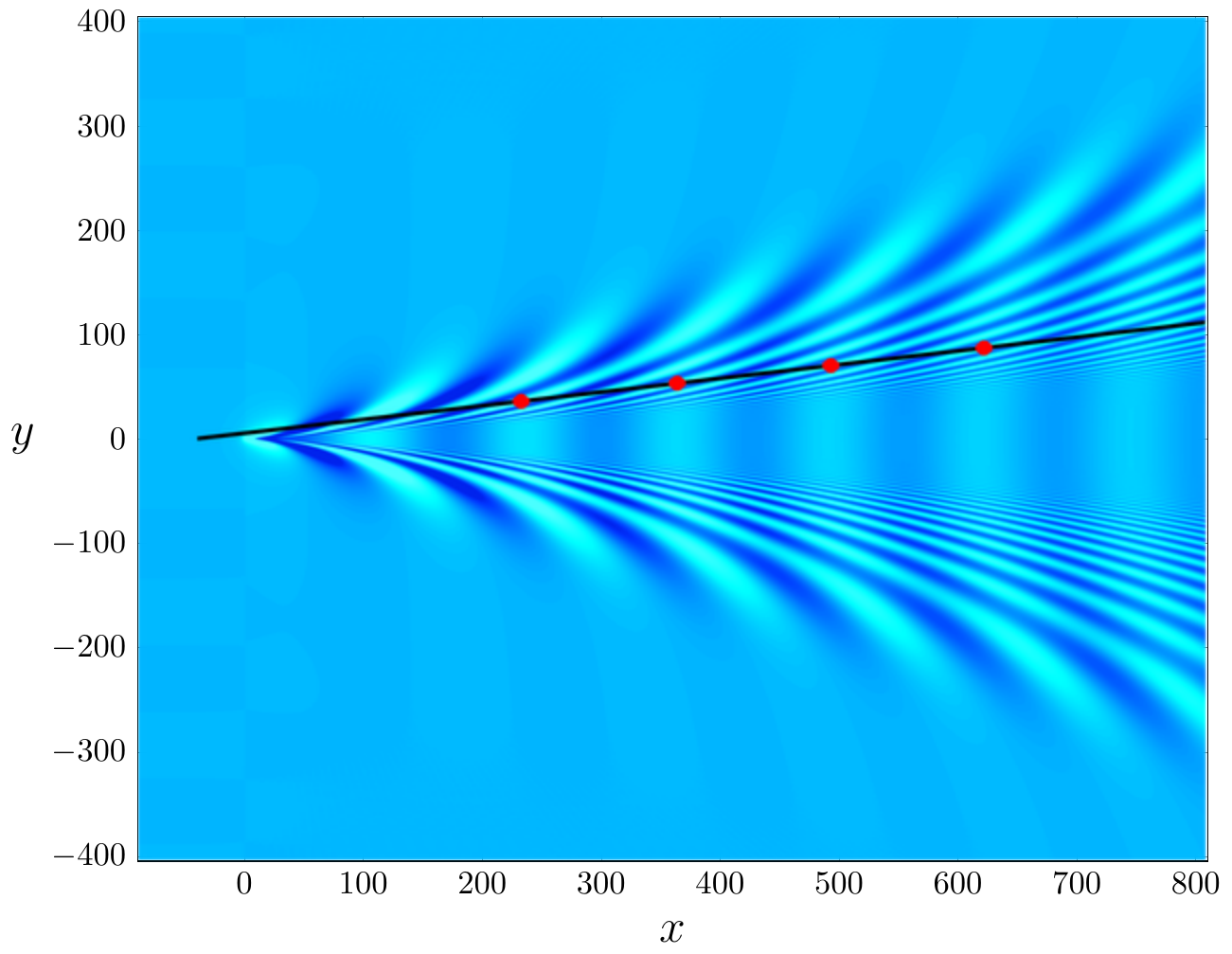}\label{fig:NonlinSurfFr45}}
\subfloat[$\epsilon=12$]{\includegraphics[width=.45\linewidth]{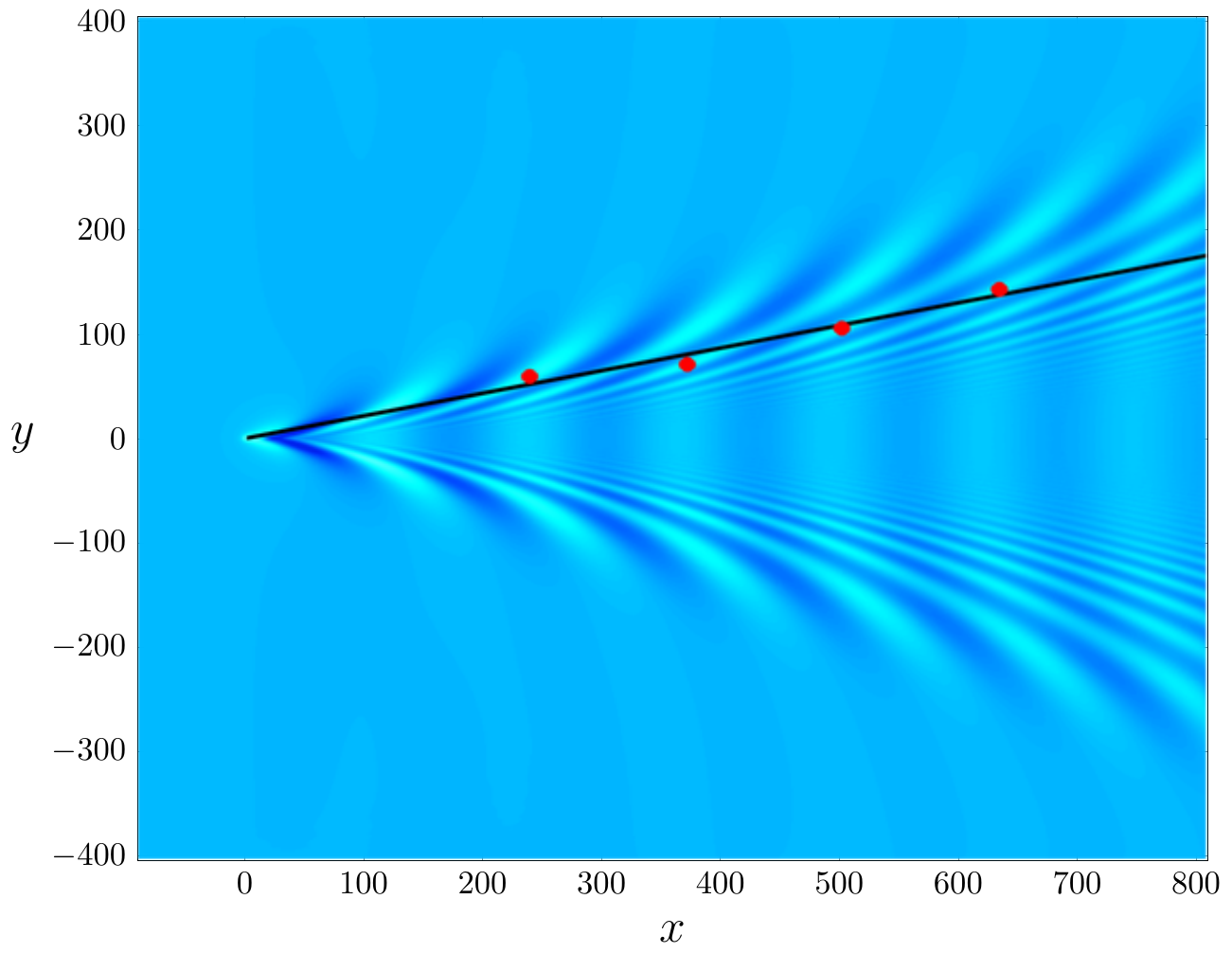}\label{fig:NonlinSurfFr45Ep12}}\\
\subfloat[$\epsilon=55$]{\includegraphics[width=.45\linewidth]{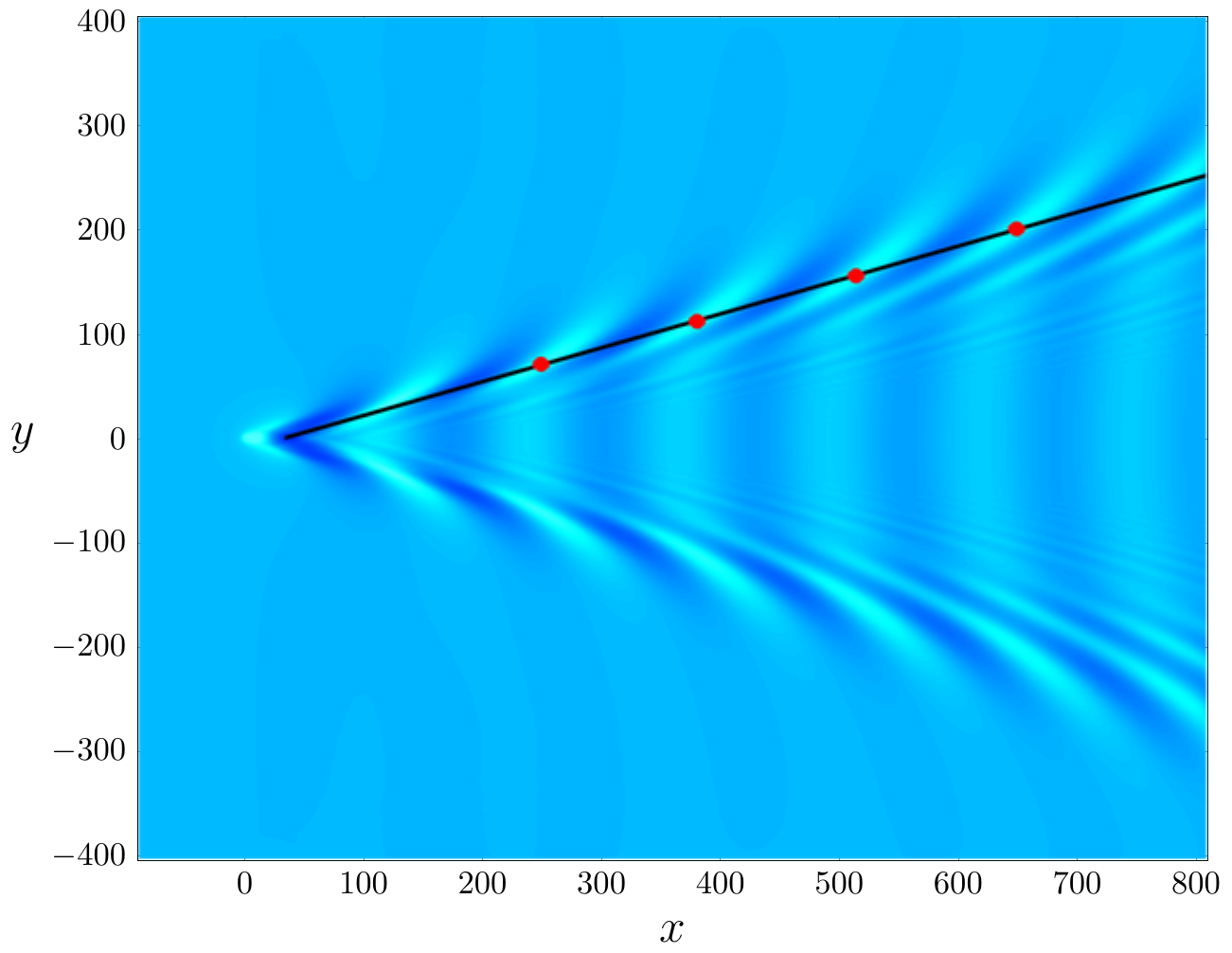}\label{fig:NonlinSurfFr45Ep55}}
\subfloat[$\epsilon=106$]{\includegraphics[width=.45\linewidth]{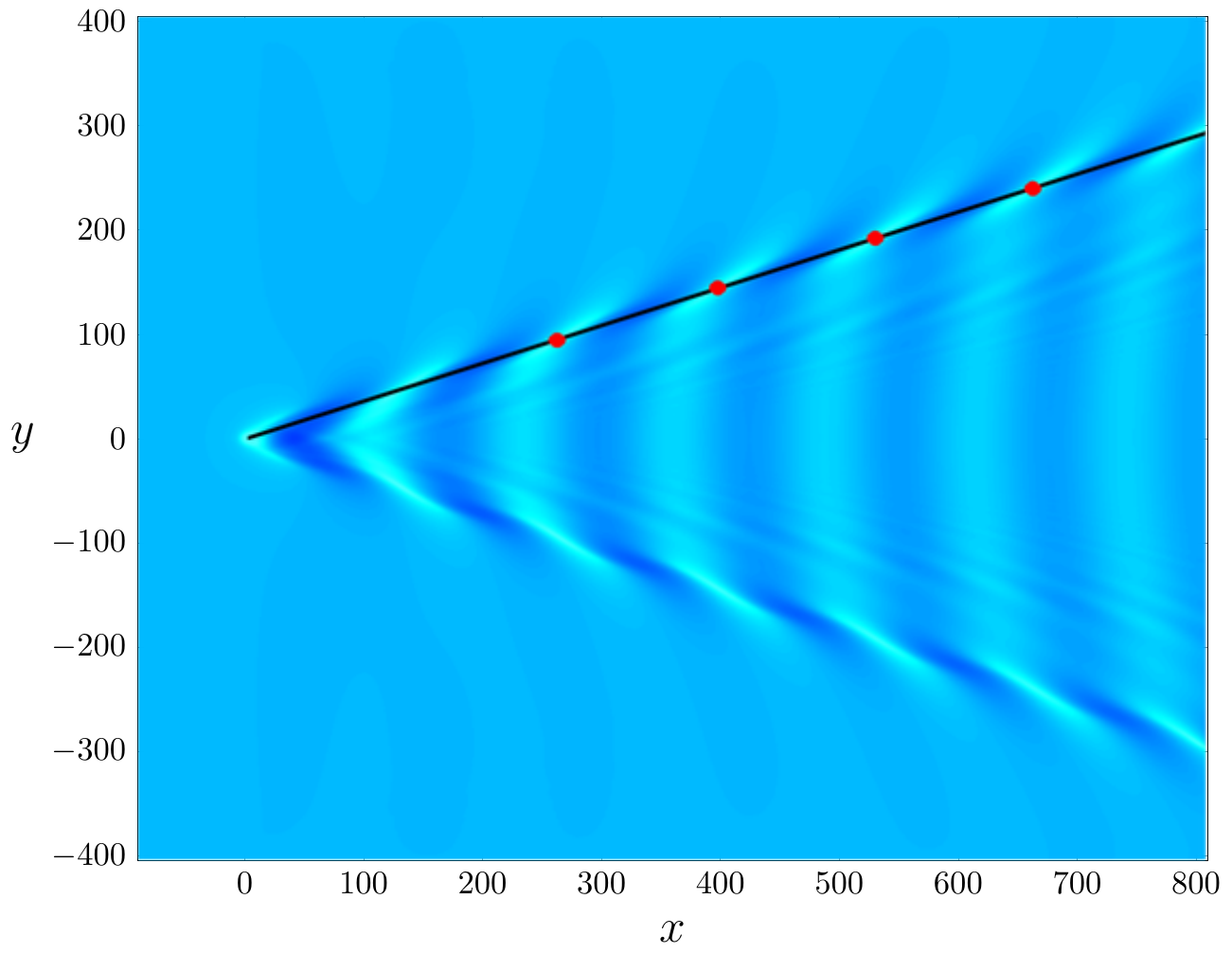}\label{fig:NonlinSurfFr45Ep106}}
\caption{Plan view of surface elevation for the problem of flow past a source with $F=4.5$: (a) the linear solution $\epsilon \ll 1$; the nonlinear solution with (b) $\epsilon=12$, (c) $\epsilon=55$ and (d) $\epsilon=106$. The (red) solid circles represent the highest peak. The (black) solid line is the line of best fit, as described in section \ref{sec:measure}.}\label{fig:NonlinSurf}
\end{figure}

\begin{figure}
\centering
\subfloat[Linear]{\includegraphics[width=.8\linewidth]{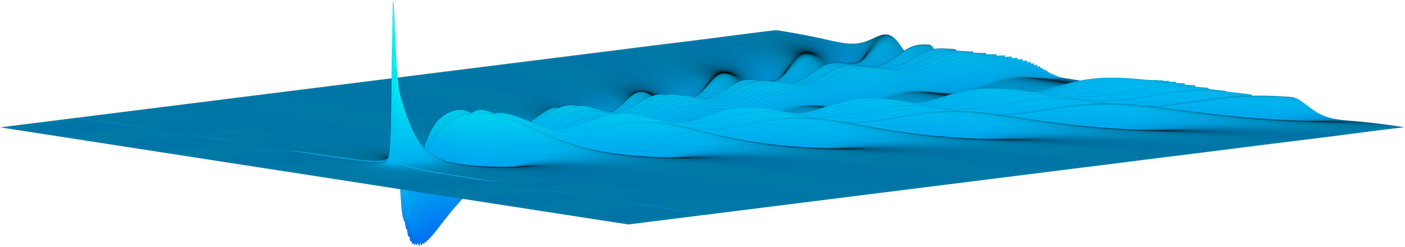}\label{fig:NonlinSurfFr45Prof}}\\
\subfloat[$\epsilon=106$]{\includegraphics[width=.8\linewidth]{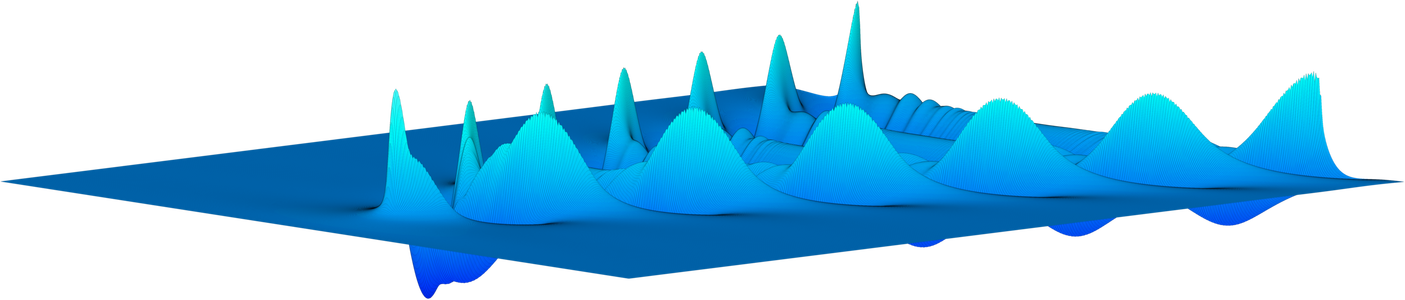}\label{fig:NonlinSurfFr45Ep106Prof}}
\caption{Free-surface profiles for flow past a source with $F=4.5$: (a) linear ($\epsilon \ll 1$) and (b) highly nonlinear ($\epsilon=106$). The $z$ direction has been scaled for comparison. A plan view of these surfaces is given in figure~\ref{fig:NonlinSurf}(a) and \ref{fig:NonlinSurf}(d).}\label{fig:NonlinSurfProf}
\end{figure}

Returning to the issue of wake angle, we have found (in almost all circumstances) that by fixing the Froude number $F$ and increasing the nonlinearity $\epsilon$ or $\mu$ (which is equivalent to increasing the size of the submerged obstacle), the apparent wake angle increases.  For example, we show in figure~\ref{fig:NonlinSurf} four free-surface patterns for flow past a source, all computed for $F=4.5$, where this trend is evident.
Further, this figure suggests the qualitative features of the wave pattern change dramatically as $\epsilon$ is varied.  These differences appear even more pronounced when viewed from an angle, as in figure~\ref{fig:NonlinSurfProf}.  We return to these figures shortly.

\begin{figure}
\centering
\subfloat[Source]{\includegraphics[width=.9\linewidth]{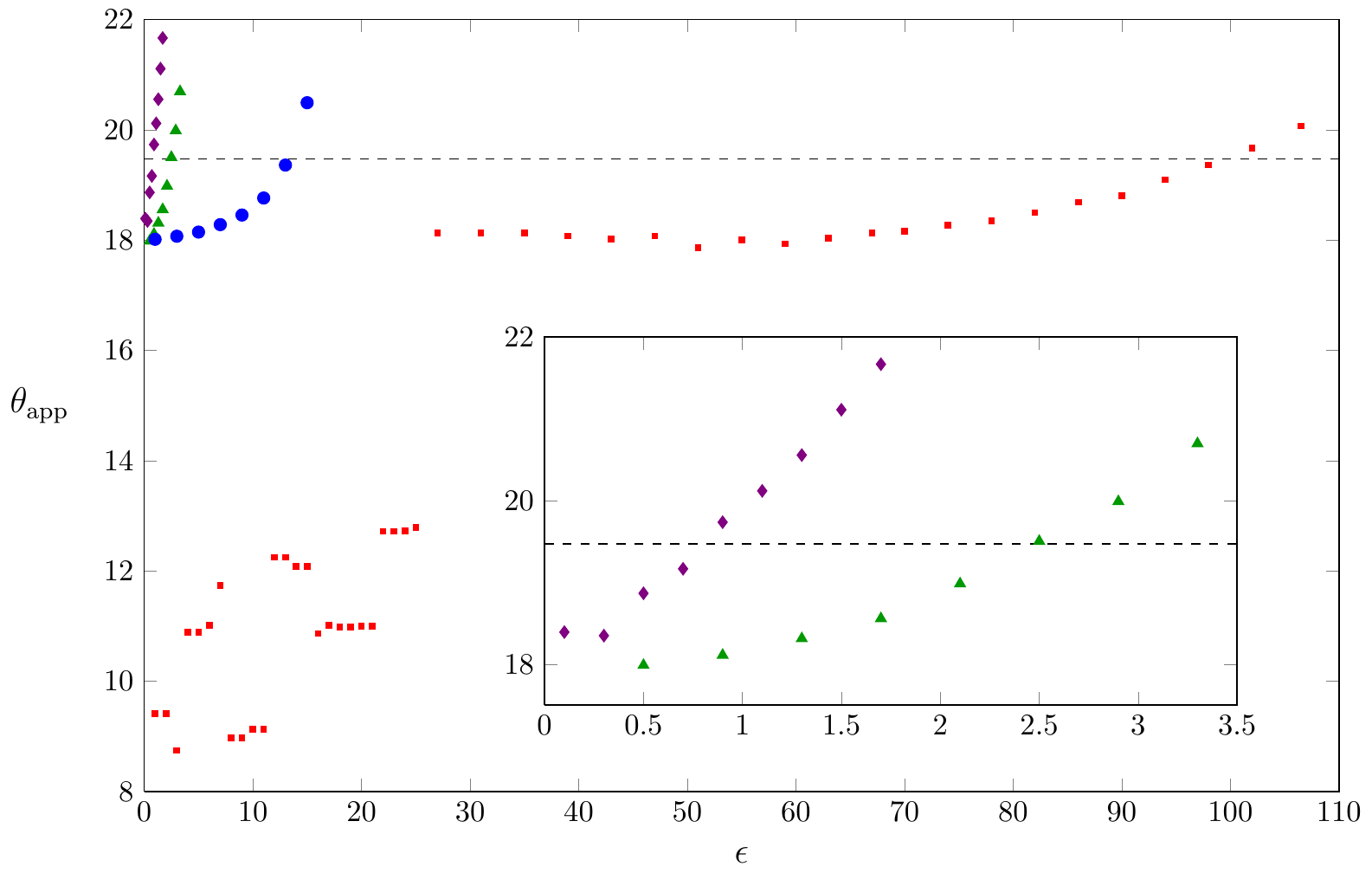}}\\
\subfloat[Dipole]{\includegraphics[width=.9\linewidth]{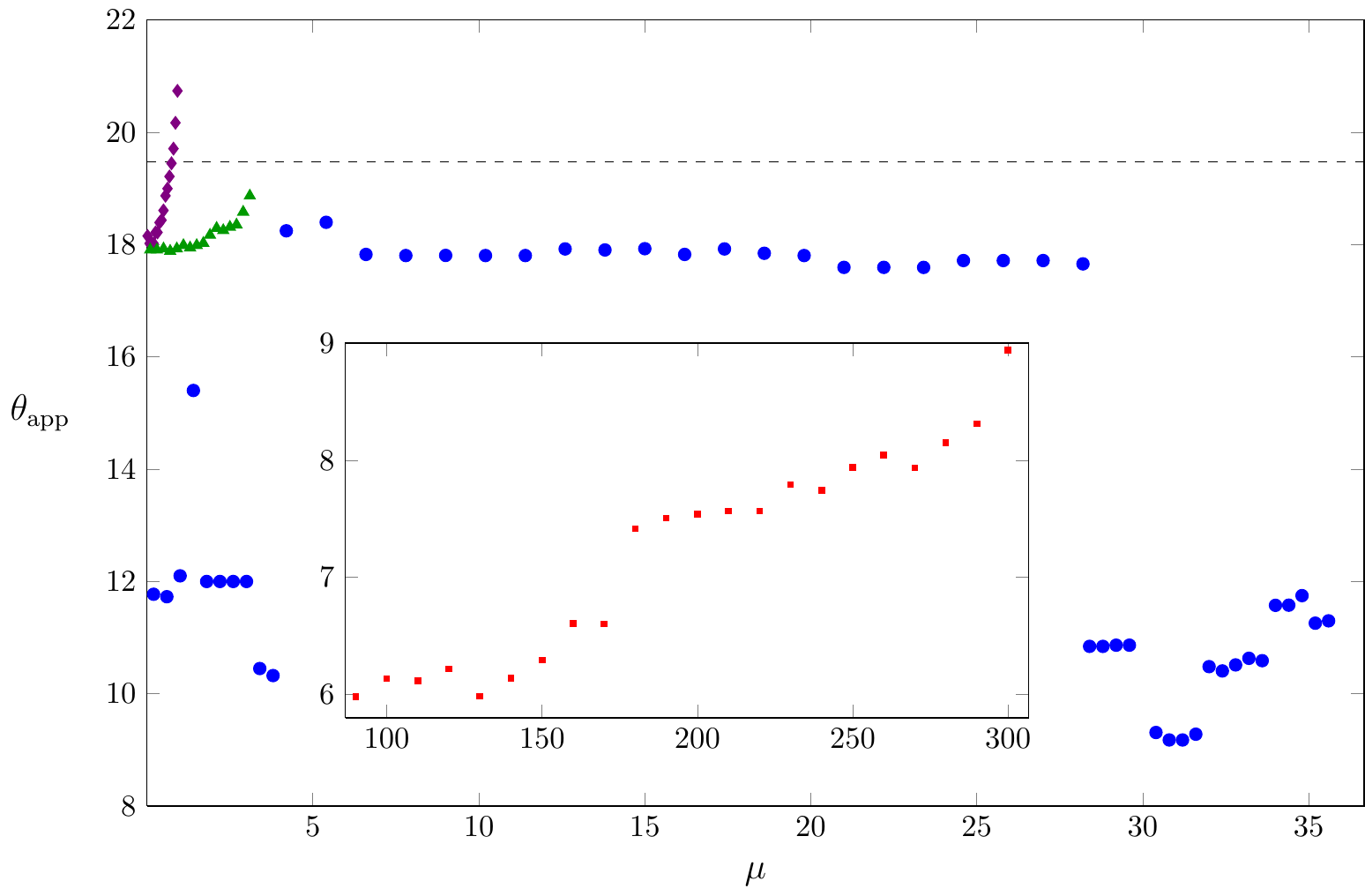}}
\caption{A plot of apparent wake angle $\theta_\mathrm{app}$ (in degrees)
against (a) $\epsilon$ for flow past a source, and (b) $\mu$ for flow past a doublet with  $F=0.9$ (violet diamonds),  $F=1.4$ (green triangles), $F=2.5$ (blue circles) and $F=4.5$ (red squares). The dashed line is Kelvin's angle, $\arcsin(1/3)\approx 19.47^\circ$. The insets provide (a) a close-up view of the results for $F=0.9$ and $1.4$, and (b) the results for $F=4.5$.
}
\label{fig:epsilonvsPeak}
\end{figure}

To demonstrate the effects of nonlinearity on apparent wake angle $\theta_\mathrm{app}$, we present in figure~\ref{fig:epsilonvsPeak} plots of $\theta_\mathrm{app}$ versus (a) $\epsilon$ for flow past a source, and (b) $\mu$ for flow past a doublet for the four Froude numbers $F=0.9$, $1.4$, $2.5$ and $4.5$.  Focusing for the moment on flow past a source (figure \ref{fig:epsilonvsPeak}(a)), we notice that for $F=0.9$, $1.4$ and $2.5$, the linear wave patterns ($\epsilon\ll 1$) fall into Regime II, with the maximum peaks lying on the outermost divergent waves and apparent wake angles which are slightly less than the Kelvin angle.  In all three cases, by increasing the measure of nonlinearity, $\epsilon$, we see a smooth increase in apparent wake angle, so that eventually it becomes greater than the Kelvin angle.  As the nonlinearity continues to increase, the solutions approach a complicated limiting configuration, as discussed above.  There are no mathematical solutions beyond this point.

Interestingly, the relationship between apparent wake angle and nonlinearity for the third case, $F=4.5$, is different to that just described for $F=0.9$, $1.4$ and $2.5$.  For $F=4.5$, the linear wave pattern ($\epsilon\ll 1$) falls into Regime III, and the nonlinear solutions for roughly $0<\epsilon<26$ also remain in this regime.  Here the apparent wake angle is much less than the Kelvin angle, as we can see by referring back to figures~\ref{fig:NonlinSurf}(a),(b), where the peaks are clearly not sitting on the outermost divergent waves.  On the other hand, at about $\epsilon = 26$ the pattern changes over to Regime II, and from that point the apparent wake angle increases smoothly with $\epsilon$ until the limiting configuration is reached.   See figures~\ref{fig:NonlinSurf}(c),(d).  The dominant features of the free-surface profiles for linear and highly nonlinear solutions are illustrated in figure~\ref{fig:NonlinSurfProf} for $F=4.5$.  In part (a), which is the linear solution, the set of highest diverging waves are found well within the Kelvin wedge.  On the other hand, in figure~\ref{fig:NonlinSurfProf}(b) we see a highly nonlinear solution for which the outermost diverging waves are shaped like dorsal fins; their amplitude is clearly much larger than the waves closer to the centreline.

\begin{figure}
\centering
\subfloat[$\epsilon=1,2$]{\label{fig:peakDiagramEp1}
\includegraphics[width=.22\linewidth]{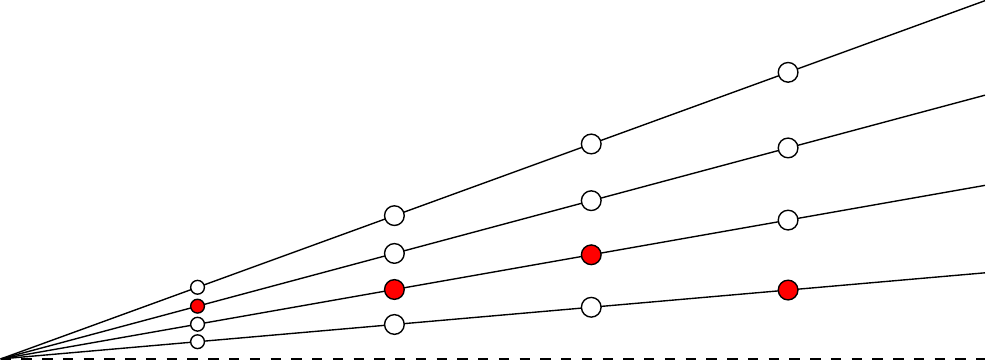}}
\subfloat[$\epsilon=3$]{\label{fig:peakDiagramEp3}
\includegraphics[width=.22\linewidth]{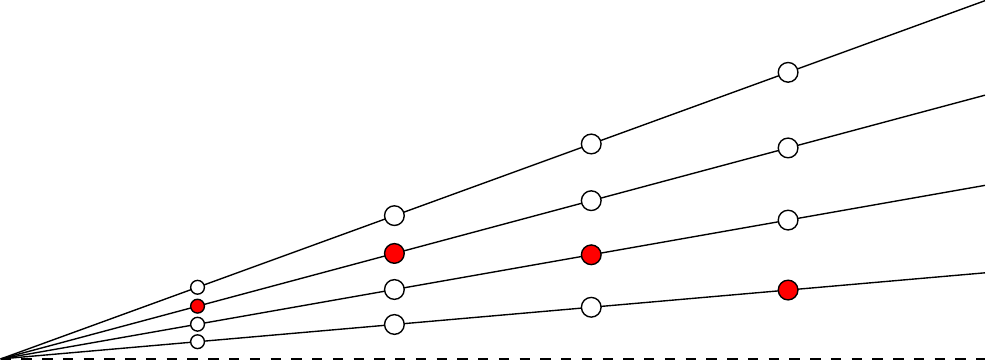}}
\subfloat[$\epsilon=4,5,6$]{\label{fig:peakDiagramEp5}
\includegraphics[width=.22\linewidth]{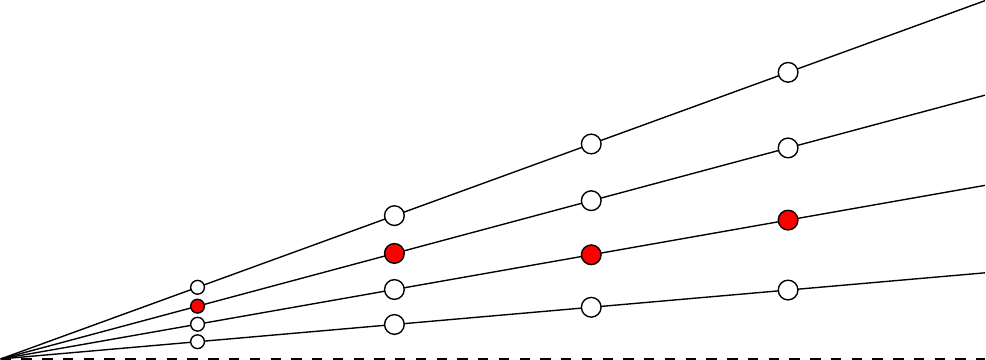}}\\
\subfloat[$\epsilon=7$]{\label{fig:peakDiagramEp7}
\includegraphics[width=.22\linewidth]{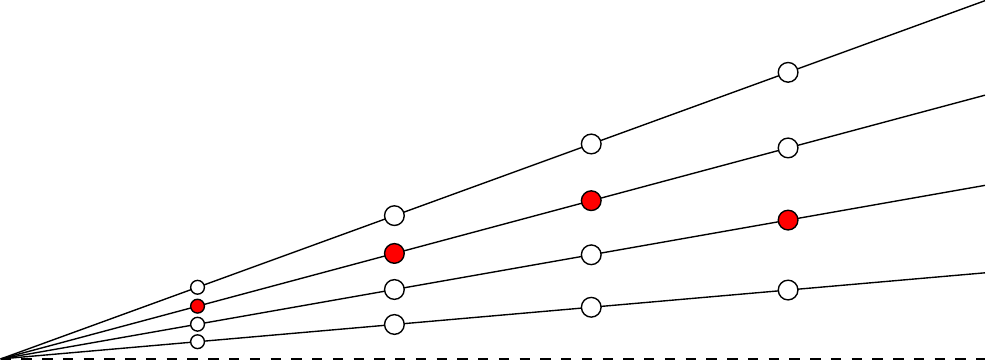}}
\subfloat[$\epsilon=8,\dots,11$]{\label{fig:peakDiagramEp10}
\includegraphics[width=.22\linewidth]{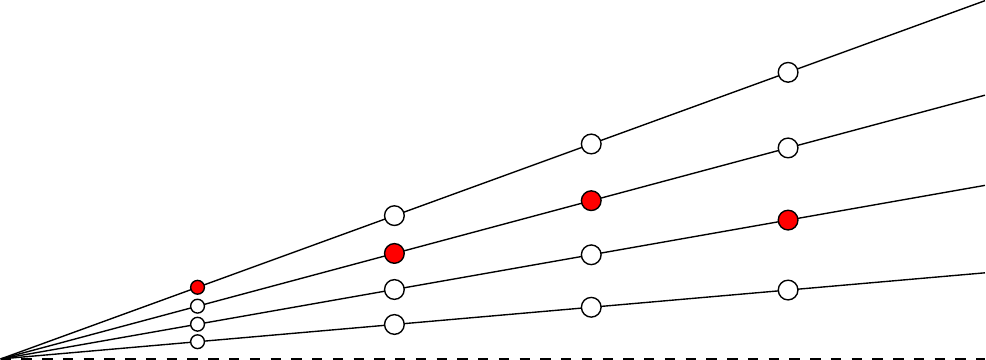}}
\subfloat[$\epsilon=12,\dots,15$]{\label{fig:peakDiagramEp13}
\includegraphics[width=.22\linewidth]{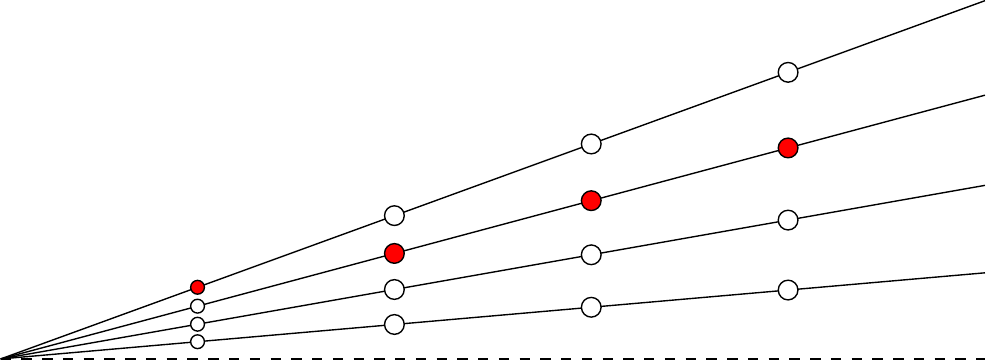}}\\
\subfloat[$\epsilon=16,\dots,21$]{\label{fig:peakDiagramEp17}
\includegraphics[width=.22\linewidth]{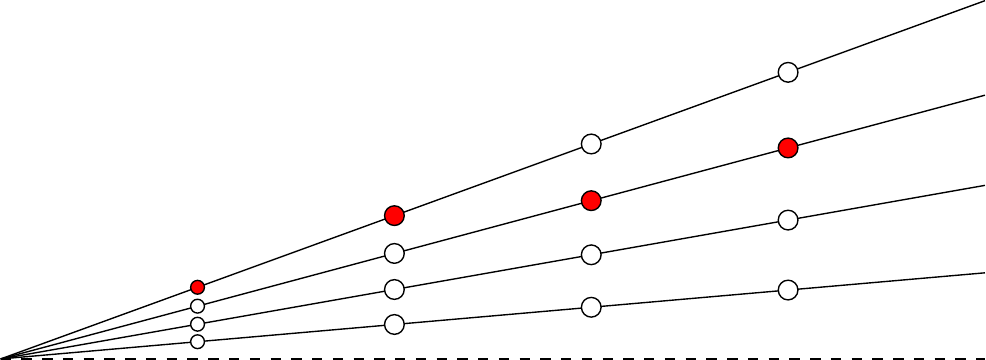}}
\subfloat[$\epsilon=22, \dots, 25$]{\label{fig:peakDiagramEp23}
\includegraphics[width=.22\linewidth]{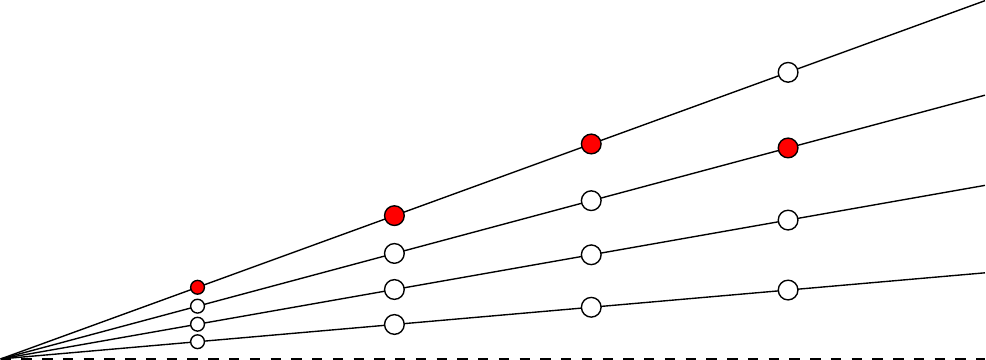}}
\subfloat[$\epsilon\geq 26$]{\label{fig:peakDiagramEp50}
\includegraphics[width=.22\linewidth]{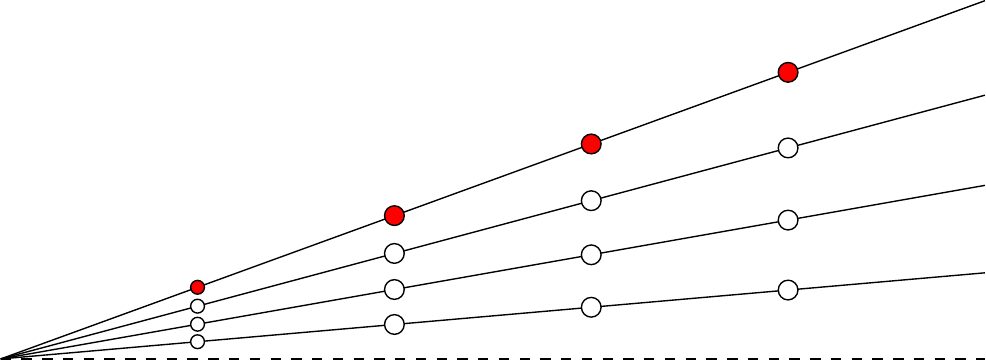}}\\
\caption{A schematic diagram indicating peaks within a given wavelength for flow past a source with $F=4.5$. The solid circles correspond to the highest peak of a given wavelength, while the open circles represent other peaks.}
\label{fig:peakDiagram}
\end{figure}

Returning to figure~\ref{fig:epsilonvsPeak}(a), the trend that the apparent wake angle $\theta_\mathrm{app}$ follows for $F=4.5$ and $\epsilon<26$ is hard to discern, as the data appears scattered.  To explain these results we have included in figure~\ref{fig:peakDiagram} a schematic diagram of which peak in a given wavelength is the highest for a given $\epsilon$ and $F=4.5$. We see that all of these solutions fall into Regime III, and as $\epsilon$ increases the highest peak within a given wavelength is located further from the centreline. While this new location of the highest peak may increase or decrease the apparent wake angle at each of these increments in $\epsilon$, there is a definite trend of an increasing wake angle, resulting in the wave pattern transitioning from Regime III to Regime II at roughly $\epsilon=26$.

Moving on to apparent wake angle for flow past a doublet, described by figure \ref{fig:epsilonvsPeak}(b), we see that for moderate Froude numbers the qualitative behaviour is similar to that for flow past a source, while there are some different behaviour for large Froude numbers.  For example, the data for $F=0.9$ and $1.4$ exhibits the same trends as for flow past a source; that is, the apparent wake angle increases with nonlinearity until the limiting configuration is reaches (when maximum peak height $\zeta_{\mathrm{max}}$ reaches the elevation $F^2/2$). For $F=2.5$, the linear solution ($\mu\ll 1$) is in Regime III, and (much like the results for flow past a source with $F=4.5$), the wave patterns transition to Regime II.  However, once in Regime II the wake angle no longer increases and instead is roughly constant for $6<\mu<28$. For $\mu>28$ the apparent wake angle transitions back into Regime III and remains there until the limiting solution is met. This example is the only case we show where the apparent wake angle does not appear to increase with nonlinearity.  Finally, while the apparent wake angles for $F=4.5$ tend to slightly increase with the measure of nonlinearity, $\mu$, the solutions remain in Regime III and do not transition to Regime II.

\begin{figure}
\centering
\subfloat[Source]{\includegraphics[width=.7\linewidth]{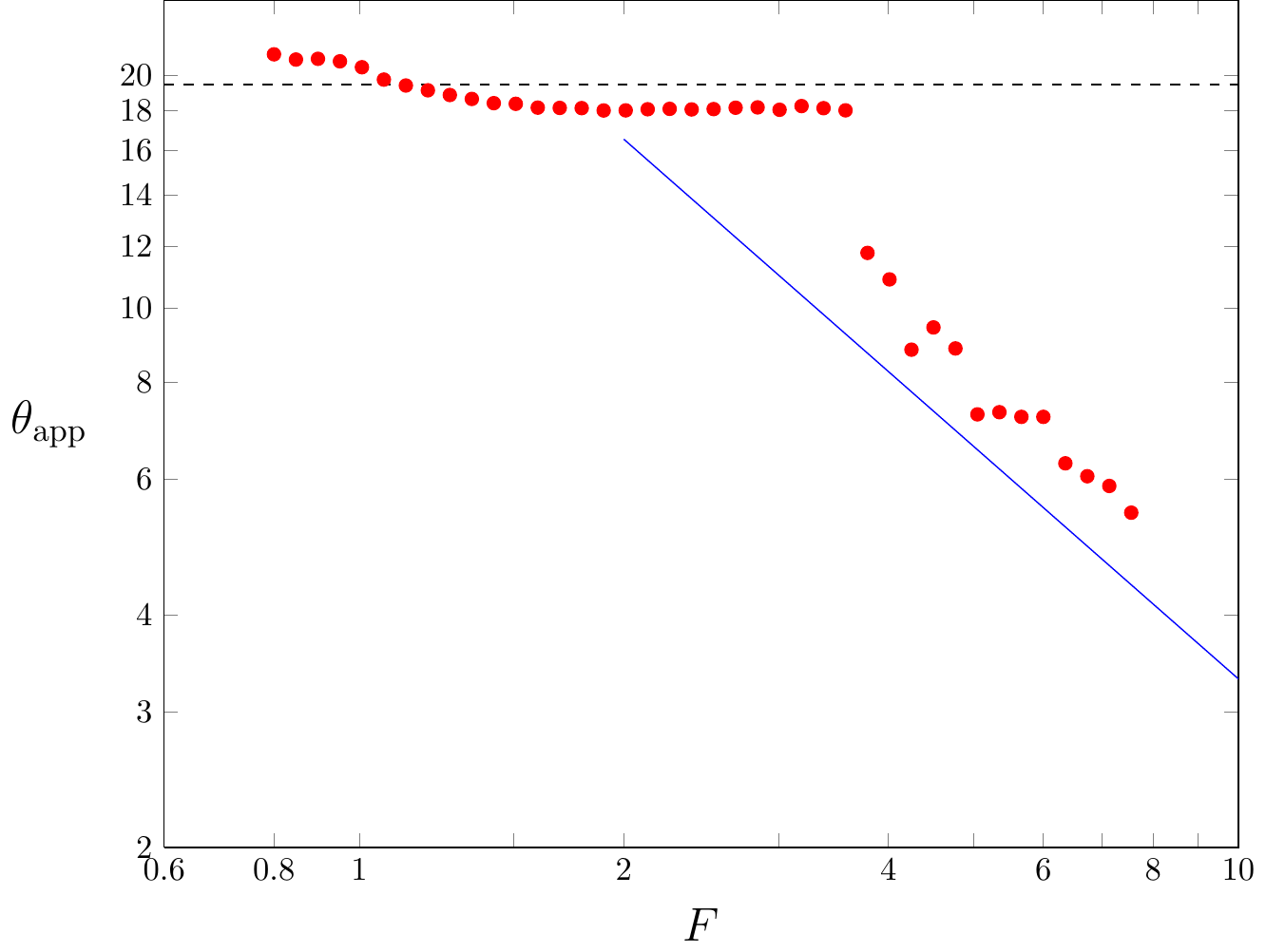}
\label{fig:FrvsPeaksource}}\\
\subfloat[Doublet]{\includegraphics[width=.7\linewidth]{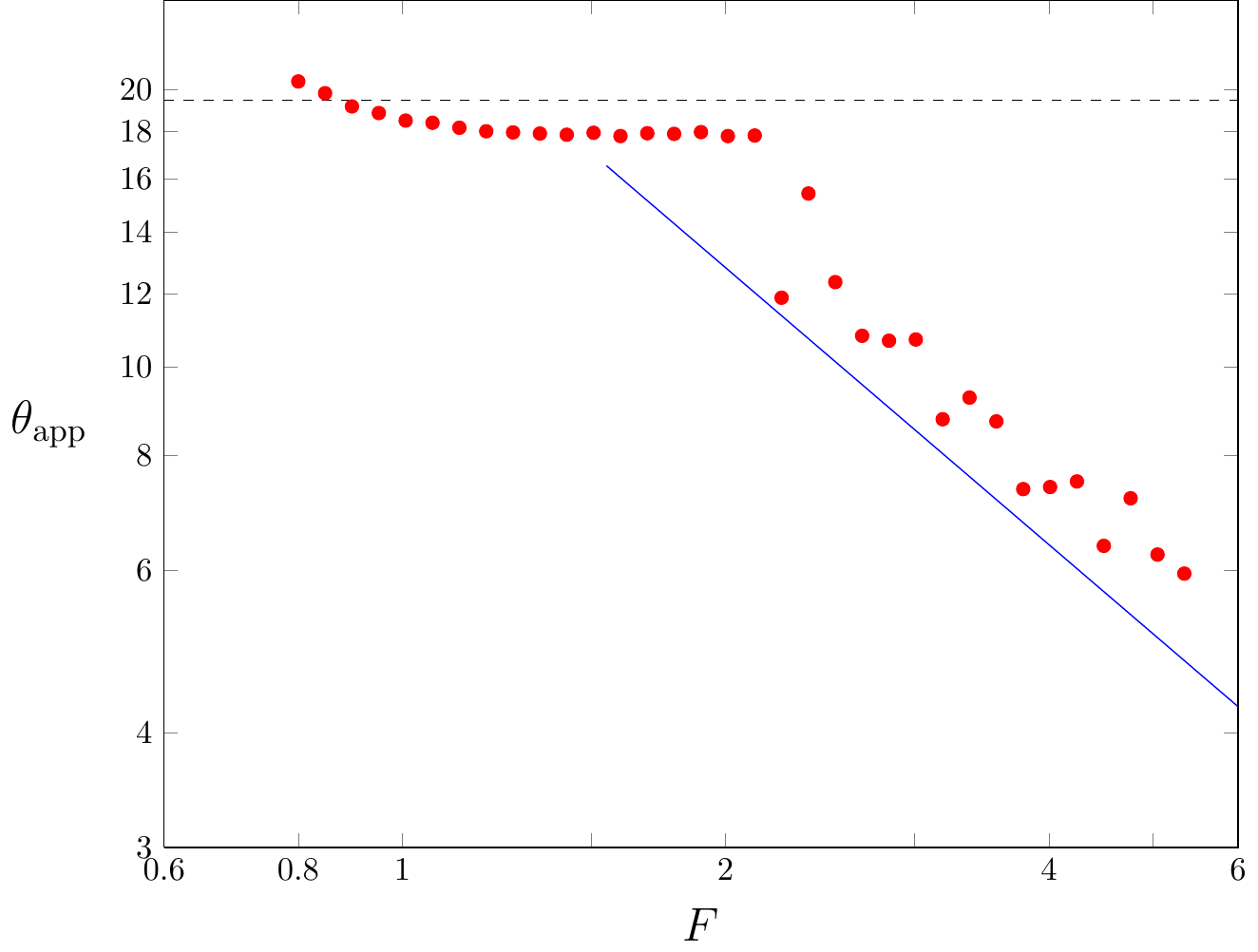}
\label{fig:FrvsPeakdoublet}}\\
\caption{Measured apparent wake angles $\theta_\mathrm{app}$ in degrees (solid circles) plotted against the Froude number for the for flow past (a) a source with $\epsilon=1.5$ and (b) a doublet with $\mu=0.7$, presented on a log-log scale. This data for nonlinear solutions is to be compared with figure~\ref{fig:FrvsPeak}, which is for linear solutions.  The dashed line is Kelvin's angle $\arcsin(1/3)\approx 19.47^\circ$.  The solid (blue) line is the theoretical asymptote shown in (a) equation (\ref{eq:AngleAsymp}) or (b) equation (\ref{eq:AngleAsympDoublet}), valid in the linear case for large Froude numbers.}
\label{fig:FrvsPeaknon}
\end{figure}

To close this section we present in figure~\ref{fig:FrvsPeaknon} two plots that aim to mimic figure~\ref{fig:FrvsPeak}, except that this time the data is from fully nonlinear calculations with $\epsilon$ and $\mu$ finite.  We have chosen to fix $\epsilon=1.5$ for figure~\ref{fig:FrvsPeaknon}(a) (flow past a source) and $\mu=0.7$ for figure~\ref{fig:FrvsPeaknon}(b) (flow past a doublet).  In both cases we observe that the general trends are the same as the linear counterparts.  For example, for moderate Froude numbers, the solutions fit in to Regime II, with the apparent wake angle close to the Kelvin angle.  For larger Froude numbers, we find the highest peaks no longer lie on the outermost divergent waves, and the wave patterns are therefore in Regime III. As with the linear solutions, the apparent wake angle for Regime III seems to scale as $F^{-1}$, although the data is a little more scattered.  The difference is that the constant of proportionality is higher for the nonlinear solutions than the linear counterpart.  To demonstrate this effect, we have included the asymptotic results (\ref{eq:AngleAsymp}) and (\ref{eq:AngleAsympDoublet}) in these log-log plots (keeping in mind they are from linear theory), and note that the nonlinear data runs roughly parallel with these lines, but the match is no longer in close quantitative agreement.

\section{Discussion}


Despite the enormous attention the theoretical study of water waves has received since the 19th century \citep{darrigol03}, there has been a resurgence of interest in ship wave patterns.  In particular, the question of what angle encloses a ship wake, or at least what angle one sees in practice,
has been the subject of very recent speculation following the insightful studies of \citet{rabaud13} and \citet{darmon14}.  A noteworthy result of this research is that for fast-moving objects, the angle formed by the highest peaks is not independent of the object's speed, but in fact decreases like the inverse Froude number.

These studies all treat mathematical solutions to linear problems, which are valid in the limit that the disturbance to the flow vanishes. In the present paper we consider the problems of free-surface flow past a submerged source and submerged doublet.  For the linearised versions of both of these problems, we also observe the decrease in apparent wake angle for sufficiently high Froude numbers.  We are able to explain our results by tracking the location of the highest peaks in a similar fashion to \citet{moisy14}, and classifying the wave patterns into three distinct regimes.  For the first problem we consider, flow past a submerged source, there is no inherent length scale in the direction of flow. Thus we eliminate the possibility of our results being influenced by interference between the disturbance's `bow' or `stern'.

We have found that nonlinear flows, for which the disturbance can no longer be considered small, provide additional interesting results.  As the nonlinearity is increased in these problems, the apparent wake angle was found to also increase.  For sufficiently fast-moving objects, there is a competition between object speed and nonlinearity, since the angle is tending to decrease with Froude number but increase with the strength of the disturbance.  Finally, we have found that for many of our highly nonlinear solutions, the apparent wake angle is greater than the Kelvin angle, so that clearly both the object speed and strength of nonlinearity are required to obtain even a rough guess of the wake angle.

\section*{Acknowledgements}
SWM acknowledges the support of the Australian Research Council via the Discovery Project DP140100933. The authors thank Prof.\ Kevin Burrage for the use of high performance computing facilities and acknowledge further computational resources and support provided by the High Performance Computing and Research Support (HPC) group at Queensland University of Technology.  We are most grateful for constructive comments and suggestions from anonymous referees.
\bibliographystyle{jfm}

\end{document}